\documentclass[aps,prb,showpacs,twocolumn,longbibliography]{revtex4-2}

\usepackage{tikz} 
\usetikzlibrary{shapes,arrows,positioning,automata,backgrounds,calc,er,patterns}
\usepackage{graphicx}
\usepackage{dcolumn}
\usepackage{bm}
\usepackage{amsmath}
\usepackage{amssymb}
\usepackage{amsthm}
\usepackage{amsfonts}
\usepackage{enumerate}
\usepackage{siunitx}
\usepackage{slashed}

\usepackage{epstopdf}

\usepackage[colorlinks=true,urlcolor=blue,citecolor=blue,linkcolor=blue]{hyperref}

\newcommand{\revision}[1]{\textcolor{blue}{#1}}

\newcommand{\affiliationSICNU}{Department of Physics, Institute of Solid State Physics and Center for Computational Sciences, Sichuan Normal University, Chengdu, Sichuan 610066, China}

\begin{document}

\title{Current response to axial gauge fields in noncentrosymmetric magnetic  Weyl semimetals}

\author{Long Liang}
\affiliation{\affiliationSICNU}

\begin{abstract}
We investigate the electric current response to axial gauge fields 
in noncentrosymmetric magnetic  Weyl semimetals. The absence of both time-reversal and inversion symmetries allows for new types of responses.
We systematically calculate the transverse, longitudinal, and Hall responses to axial gauge potentials with both linear and quadratic dispersion relations. The transverse and Hall responses are of  comparable magnitude, while the longitudinal response is much smaller. Notably, with increasing  frequency, the transverse and Hall response functions manifest a peak whose height is determined by the properties of Weyl fermions and is independent of the axial gauge potential. The main features of the response functions survive in the presence of disorders.
As applications of our results, we propose a Hall type magnetopiezoelectric effect, where a transverse sound wave can induce an electric current whose direction is perpendicular to the directions of sound propagation and polarization. Our results also provide a mechanism to excite magnons  using electric fields and could be useful for magnon spintronics.
\end{abstract}
\maketitle

{\it Introduction.\textemdash}
In recent years, there has  a surge of interest in topological 
quantum matter with gapless nodes and linear dispersion relation around the nodes. Typical materials include Dirac and Weyl semimetals and their generalizations~\cite{Armitage:rev-2018,HongDingRMP}. 
The nodal degrees of freedom in these materials allow for the realization of responses that go beyond the conventional light-matter interaction paradigm~\cite{Ilan-Pikulin:rev-2019,Yu_2021}. 
In particular, axial gauge fields, which are out of phase motion of the nodes, can be manipulated through various methods such as optical~\cite{PhysRevB.102.245123}, electrical~\cite{PhysRevLett.125.266601}, magnetic, and mechanical~\cite{PhysRevB.94.214306,Yuan2020,PhysRevB.94.241405,pub.1142612917}. Indeed,  besides the response to electromagnetic  fields~\cite{PhysRevB.85.165110,PhysRevLett.109.181602,Zhou_2013,PhysRevB.87.245131,PhysRevB.93.085442,PhysRevB.94.165111,PhysRevB.96.155117,PhysRevB.102.235134,PhysRevB.102.045148,PhysRevB.108.035407,PhysRevB.97.035144}, the response of topological semimetals to axial gauge fields has also attracted much attention and various novel phenomena~\cite{Cortijo-Vozmediano:2016,Pikulin-Franz:2016,Grushin:2016,Landsteiner_2016,PhysRevB.96.085201,PhysRevX.7.041026,PhysRevB.95.115410,PhysRevLett.124.126602,LLTO2020,PhysRevB.102.121105,PhysRevB.102.241401,PhysRevLett.124.126602,AMEE,PhysRevB.106.045132,ACME} have been proposed.

Noncentrosymmetric magnetic Weyl semimetals that break both symmetries have been proposed in a number of materials such as Huesler magnets like Ti$_2$Mn$X$ ($X$=Al, Ga, or In)~\cite{YanSun1,YanSun2} and InMnTi$_2$~\cite{Heusler}, and the rare earth family of compounds $R$Al$X$  where $R$=rare earth and $X$=Ge or Si~\cite{PhysRevB.97.041104}. 

Recently, several candidate materials in the rare earth family have been experimentally identified~\cite{Sanchez2020,Gaudet2021,PhysRevX.13.011035,Drucker2023,PhysRevB.109.035120,PhysRevB.109.195130}. Motivated by these experimental discoveries, in this work, we study the electric current response to axial gauge fields in noncerntrosymmetric magnetic Weyl semimetals, where the absence of both time-reversal and inversion symmetries paves the way for new types of responses, which are otherwise forbidden in Weyl semimetals break only inversion or time-reversal symmetry. In particular, to generate electric currents, the external perturbations need not break inversion nor time-reversal. 

We systematically calculate the diagonal longitudinal and transverse response functions as well as the off-diagonal Hall type response function. We consider both linear  (e.g., sound wave) and quadratic dispersion (e.g., ferromagnetic spin wave) relations of external perturbations. The frequency dependence of the response functions is obtained. We find that the transverse and Hall type responses are of comparable magnitude, while the longitudinal one is substantially smaller. Moreover, the momentum dependence of the axial gauge potential results in a peak in the response functions. For the transverse and Hall responses, the height of the peak is independent of external perturbations. 
As applications of our results, we predict a Hall type magnetopiezoelectric effect~\cite{PhysRevLett.117.257601,PhysRevB.96.064432,PhysRevB.97.235128,PhysRevLett.122.127207},
and also propose a possible mechanism of magnon electric field interconversion.

{\it Model.\textemdash}
To  elucidate the essential features of the  response to axial gauge potentials, we employ a minimal two band model of a Weyl semimetal that breaks both time-reversal and inversion symmetries. The corresponding Hamiltonian is $H=H_{+}\oplus H_{-}$ with
\begin{eqnarray}
	H_{\lambda}=\hbar v_F \lambda (\mathbf{k}-\lambda \mathbf{b})\cdot\bm{\sigma} + \lambda b_0 \sigma^0,
\end{eqnarray}
where $\lambda=\pm 1$ denotes the chirality of the Weyl nodes, $\hbar$ is the reduced Planck's constant, $v_F$ is the Fermi velocity,  $\sigma_i$ with $i=1,2,3$ are the three Pauli matrices, $\sigma^0$ is the two dimensional identity matrix, $2\mathbf{b}$ and $2b_0$ are the separations of the two Weyl nodes in momentum and energy, respectively. The inversion symmetry is broken by nonzero $b_0$,  while $\mathbf{b}\ne 0$ breaks the time-reversal symmetry. In the following, the Fermi velocity, reduced Planck's constant, Boltzmann constant, and electric charge are taken to be unity and will be recovered when necessary.

In the low energy limit, the effects of external perturbations can be captured by axial gauge potentials~\cite{Ilan-Pikulin:rev-2019,Yu_2021}. This provides a general way to investigate the current response to external perturbations. 
The electric current induced by the scalar axial gauge potential has been explored previously~\cite{ACME}. We thus focus on the effect of vector axial gauge potential $\mathbf{A}_{5}$.
Using the linear response theory, the electric current induced by the axial vector potential is given by 
	$j_i(\omega,\mathbf{p})=\Pi_{ij}(\omega,\mathbf{p})A_{5,j}(\omega,\mathbf{p})$.
We calculate the retarded current\textendash axial current response function $\Pi_{ij}(\omega,\mathbf{p})$  by utilizing the imaginary time formalism,
	$\Pi_{ij}(i\nu_m,\mathbf{p})=\frac{1}{\beta}\sum_{k}\mathrm{tr}[\hat{j}_i G(k+q) \hat{j}_j\gamma_5 G(k)]$,
where $\beta$ is the inverse temperature, $k\equiv (i\omega_n,\mathbf{k})$, $\omega_n=(2n+1)\pi/\beta$ is the fermionic Matsubara frequency, and $\nu_m=2m\pi/\beta$ is the bosonic one.
The current operator is given by $\hat{j}_i=\partial_{k_i}H(\mathbf{k+q}/2)=\sigma_i\oplus -\sigma_i$, and $\gamma_5=I\oplus -I$ is the fifth gamma matrix.
The  Matsubara Green's function is given by 
	$G(i\omega_n,\mathbf{k})=[i\omega_n-H(\mathbf{k})-\mu]^{-1}$,
with $\mu$ being the chemical potential. In the following we assume $\mu>0$.

{\it Response function.\textemdash}
Now we calculate the response function. Performing Matsubara frequency summation and analytical continuation, $i\nu_m \to \omega+i0^+$, we obtain  the retarded response function $\Pi_{ij}(\omega,\mathbf{p})$,
\begin{eqnarray}
	\Pi_{ij}(\omega,\mathbf{p})
	=\sum_{\mathbf{k},\lambda,s,s'}
	\frac{N^{\lambda ss'}_{ij}(\mathbf{k},\mathbf{p})[f(\epsilon^{\lambda}_{s,\mathbf{k+p}})-f(\epsilon^{\lambda}_{s',\mathbf{k}})]}{\omega+\epsilon^{\lambda}_{s',\mathbf{k}}-\epsilon^{\lambda}_{s,\mathbf{k+p}} +i0^+},
\end{eqnarray}
where $\epsilon^\lambda_{s,\mathbf{k}}=s |\mathbf{k}-\lambda \mathbf{b}|+\lambda b_0$ with $s=\pm 1$ denotes the particle and hole bands, $f(\epsilon)=1/(e^{\beta(\epsilon-\mu)}+1)$ is the Fermi-Dirac distribution function. In this work we focus on the zero temperature limit. The transition amplitude $N^{\lambda ss'}_{ij}(\mathbf{k},\mathbf{p})$ is given by
\begin{widetext}
	\begin{eqnarray}\label{Eq:N}
		N^{\lambda ss'}_{ij}(\mathbf{k},\mathbf{p})
		=\frac{1}{2}\left[\bigg(1-ss'\frac{\mathbf{k}^\lambda \cdot(\mathbf{k}^\lambda +\mathbf{p})}{|\mathbf{k^\lambda }||\mathbf{k^\lambda +p}|}\bigg)\delta_{ij}\right.
		+ss'\frac{k^\lambda_i(k^\lambda_j+p_j)+k^\lambda_j(k^\lambda_i+p_i)}{|\mathbf{k^\lambda}||\mathbf{k^\lambda +p}|}
		\left.+i\epsilon_{ijl}\bigg(s'\frac{k^\lambda_ l}{|\mathbf{k}^\lambda|}-s\frac{k^\lambda_l+p_l}{|\mathbf{k^\lambda+p}|}\bigg)
		\right],
	\end{eqnarray}
\end{widetext}
with $\mathbf{k}^\lambda=\mathbf{k}-\lambda \mathbf{b}$ and $\epsilon_{ijl}$ being the totally antisymmetric tensor. The $s=s'$ and $s\ne s'$ terms correspond to intraband and interband transitions, respectively. 

The first two terms in Eq.~\eqref{Eq:N} are symmetric in $i$ and $j$, while the third term is antisymmetric. The response function is a second order tensor constructed from  $\mathbf{p}$ and  $\epsilon_{ijk}$, so it must take the form
$\Pi_{ij}(\omega,\mathbf{p})=\Pi_T(\omega,p)(\delta_{ij}-p_ip_j/p^2)+\Pi_L(\omega,p)p_ip_j/p^2+\Pi_H(\omega,p)\epsilon_{ijk}p_k/p$.
Here $\Pi_T(\omega,p)$ is the transverse response function, $\Pi_L(\omega,p)$ is the longitudinal one, and $\Pi_H(\omega,p)$ describes a Hall-type response.

{\it Longitudinal and transverse responses.\textemdash}
We first consider the longitudinal and transverse response functions.
It is convenient to first calculate their imaginary parts, 
\begin{widetext}
\begin{eqnarray}
	&&\mathrm{Im}\Pi^{\mathrm{intra}}_{L/T}(\omega,p)
	=\sum_\lambda\lambda\theta(p-|\omega|)\theta(2|\mu_\lambda|+\omega-p)g_{L/T}F_{L/T}(-\omega,|\mu_\lambda|,p)-(\omega\to-\omega),\label{Eq:ImPiDIntra}\\
		&&\mathrm{Im}\Pi^{\mathrm{inter}}_{L/T}(\omega,p)
	=\sum_\lambda\lambda\theta(|\omega|-p) 
	g_{L/T}\bigg[\theta(p-|\omega-2\mu_\lambda|)F_{L/T}(\omega,\mu_\lambda,p)
	+\frac{\theta(\omega-2\mu_\lambda-p)}{24\pi } 
	\bigg]-(\omega\to -\omega),\label{Eq:ImPiDInter}
\end{eqnarray}
\end{widetext}
where $\theta(x)$ is the step function,  $\mu_\lambda=\mu+\lambda b_0$, $g_L=\omega^2$, $g_T=\omega^2-p^2$, $F_{L}(\omega,\mu,p)=\omega^2 (2\mu-\omega-p)^2(2\mu+2p-\omega)/(96 \pi p^3)$, 
 and 
$F_{T}=(\omega+p-2\mu)[(2\mu-\omega)(2\mu-\omega+p)+4p^2]/(192\pi p^3)$. It is clear that if $|\omega|<v_Fp$ (here we recover the Fermi velocity), only the intraband term is nonzero, while the interband processes contribute in the opposite limit. 
The imaginary parts of the longitudinal and transverse response functions are odd in frequency, which implies that their real counterparts are even functions. In the following we only consider the $\omega>0$ region unless otherwise stated.
\begin{figure}[h]
	\includegraphics[width=0.43\textwidth]{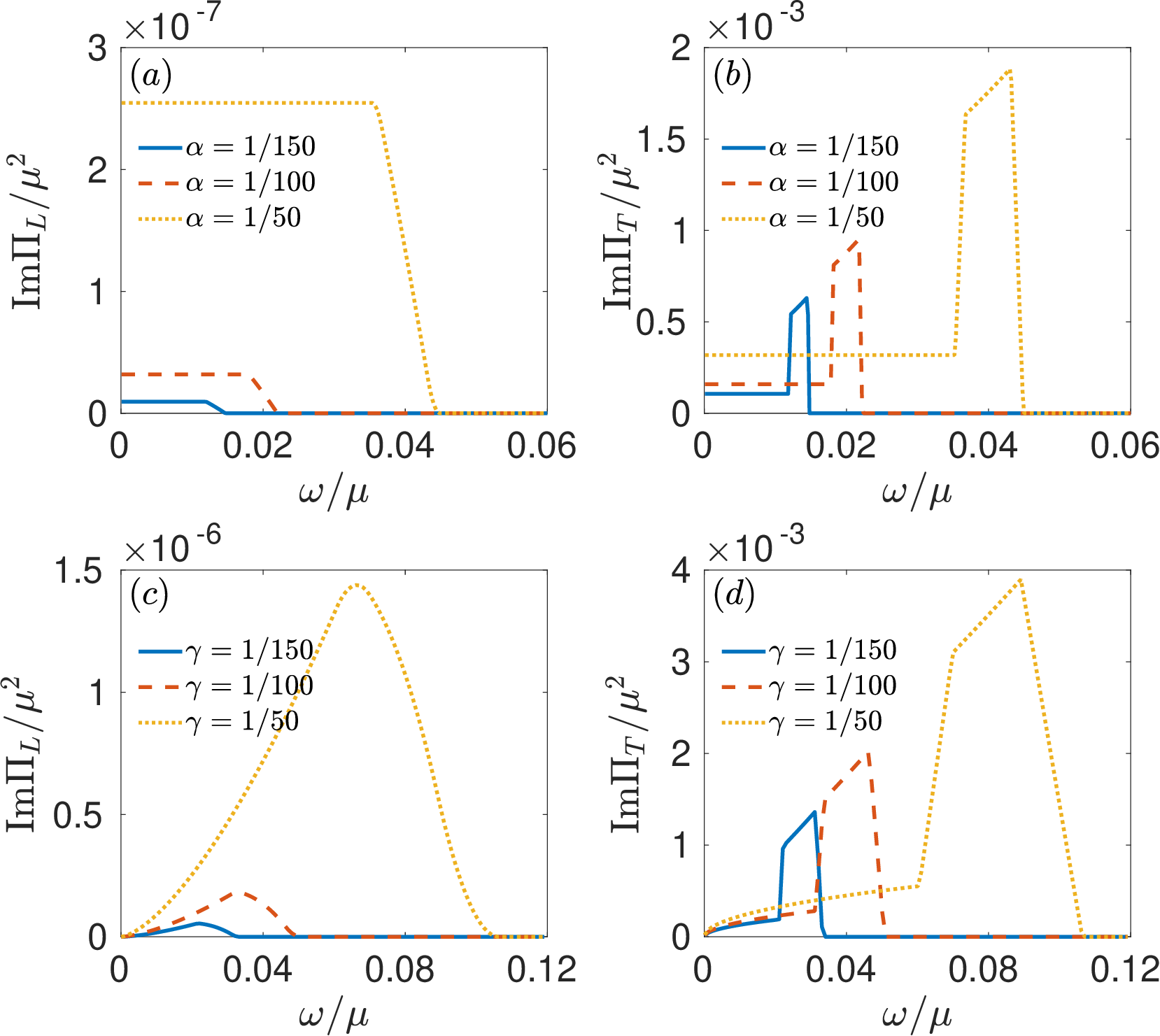}
	\caption{The imaginary parts of the longitudinal and transverse response functions for $\omega=v_sp\equiv \alpha v_F p$ [(a)-(b)] and $\omega=Dp^2\equiv \gamma v_F/k_F p^2$ [(c)-(d)]. Here we take the chemical $\mu$ and Fermi velocity $v_F$ to be unity and $b_0$ is fixed to be $0.1\mu$.}\label{Fig:Pi_d}
\end{figure}

The dispersion relation of the axial gauge field depends on  its physical origin. For example, for sound waves and antiferromagnetic spin waves, $\omega=v_s p$ in the low energy limit, where $v_s$ is the sound or spin wave velocity. The low energy dispersion relation for ferromagnetic spin waves is $\omega=D p^2$, with $D$ being the spin stiffness. For helical magnetic order observed in Weyl semimetals~\cite{Gaudet2021,PhysRevX.13.011035,Drucker2023}, the magnon dispersion is highly anisotropic~\cite{PhysRevB.73.054431}.
In real materials, we typically have $\alpha\equiv v_s/v_F\ll 1$ and $\gamma\equiv Dk_F/v_F\ll 1$ ($k_F$ is the Fermi momentum)~\footnote{The Fermi velocity $v_F$ of Weyl semimetals  is of the order of $10^6$\si{m/s}, while the sound velocity $v_s$ is of the order of $10^3$\si{m/s}, so $\alpha =v_s/v_F\ll 1$. The spin stiffness $D$ is typically of the order of \si{meV\AA}$^2$, while $\hbar v_F$ is of the order of \si{eV\AA} and the Fermi momentum $k_F$ is of the order of \si{\AA}$^{-1}$, thus $\gamma=D k_F/v_F\ll1$.}. 
In the following we study both the linear and quadratic dispersions of the axial gauge potential and focus on the $\alpha \ll 1$ and $\gamma\ll 1$ regimes.
Fig.~\ref{Fig:Pi_d} shows the imaginary parts of the longitudinal and transverse responses functions for both linear and quadratic dispersions as functions of frequency. The step-like feature can be observed, and one can also see that the transverse response is much larger than the longitudinal one. In the linear dispersion case [Fig.~\ref{Fig:Pi_d}(a) and (b)], the interband term~Eq.~\eqref{Eq:ImPiDInter} vanishes identically since $v_F/v_s>1$. For the quadratic dispersion case [Fig.~\ref{Fig:Pi_d}(c) and (d)],  the interband contribution also vanishes in the plotted frequency region.

\begin{figure}
	\includegraphics[width=.43\textwidth]{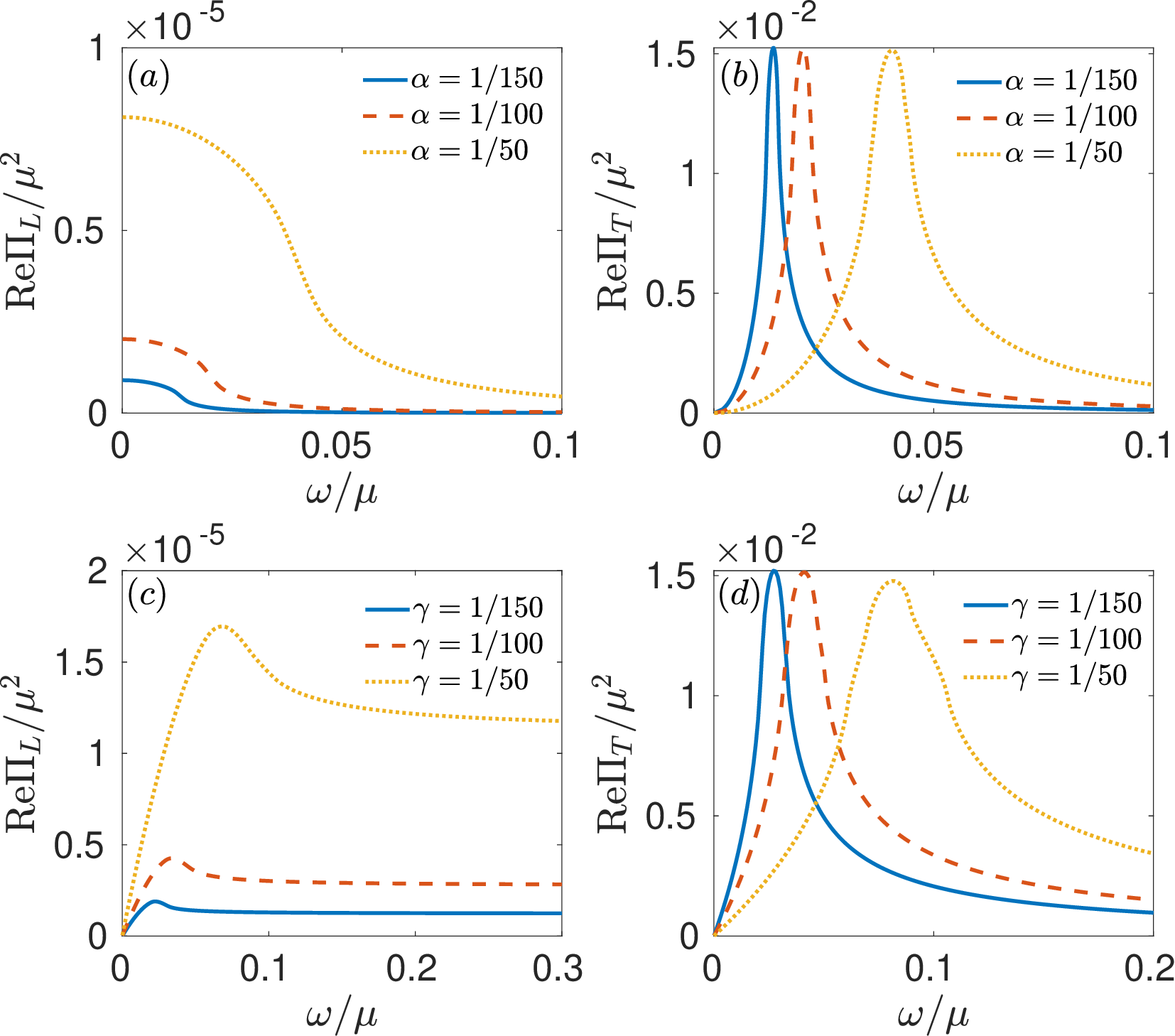}
	\caption{The real parts of the longitudinal  and transverse response functions for linear [(a)-(b)] and quadratic  [(c)-(d)] dispersion relations. Other parameters are the same as in Fig.~\ref{Fig:Pi_d}. The peaks in (b) and (d) are located at $\omega\sim 2\alpha\mu$ and $\omega\sim 4\gamma\mu$, respectively.}\label{Fig:Pi_d_Re}
\end{figure}

The real parts of the response functions are obtained through the Kramers-Kronig relation from the imaginary parts.
Fig.~\ref{Fig:Pi_d_Re} shows the real parts of the response functions with the same parameters as in Fig.~\ref{Fig:Pi_d}.
For the linear dispersion case, the real part of the longitudinal response function decreases  monotonically with the increasing of frequency, see Fig.~\ref{Fig:Pi_d_Re}(a). In the $b_0/\mu\ll1$  and  $\omega/\mu\ll 1$ limit, we find
\begin{eqnarray}
	\mathrm{Re}\Pi_{L}(\omega,p)\approx \frac{2\mu b_0 \alpha^2}{\pi^2}
	\left(1-\frac{\omega^2}{12\mu^2 \alpha^2}\right),
\end{eqnarray}
which is nonvanishing in the zero frequency limit. In the large $\omega$ limit, 
\begin{eqnarray}\label{Eq:PiL_large}
	\mathrm{Re}\Pi_{L}(\omega,p)\approx \frac{8\mu b_0(\mu^2+b^2_0) \alpha^4}{3\pi^2}\frac{1}{\omega^2},
\end{eqnarray}
which decreases as $1/\omega^2$ and is suppressed by $v_F^4/v^4_s$.

In contrast to the longitudinal counterpart, the real part of the transverse response function first increases with increasing $\omega$  and then decreases after reaching a maximum, see  Fig.~\ref{Fig:Pi_d_Re}(b). 
In the small $\omega$ and $b_0$ limit, 
\begin{eqnarray}\label{Eq:PiT_small}
	\mathrm{Re}\Pi_T(\omega,p)
	\approx-\frac{2\mu b_0 \alpha^2}{\pi^2}\left(1-\frac{\omega^2}{12\mu^2 \alpha^4}\right),
\end{eqnarray}
which is also nonzero in the $\omega\to 0$ limit but takes the opposite value compared to the longitudinal response. 
One may expect that, in the long wave length limit, the electrons cannot distinguish between the transverse and longitudinal waves, resulting in  identical responses. This holds true if the frequency remains finite as the momentum approaches to zero; in our situation, however, both $\omega$ and $p$ approach to zero, and  since $v_F p \gg \omega$, the momentum transfer cannot be ignored and the system responses to longitudinal and transverse waves in different manners even in the zero frequency and momentum limit. 

In the large $\omega$ limit, the real part of the transverse response is
\begin{eqnarray}\label{Eq:PiT_large}
	\mathrm{Re}\Pi_{T}(\omega,p)\approx \frac{8\mu b_0(\mu^2+b^2_0)\alpha^2 }{3\pi^2}\frac{1}{\omega^2},
\end{eqnarray}
which deceases as $1/\omega^2$ and is suppressed by $v^2_F/v^2_s$. From the low and high frequency expansions Eqs.~\eqref{Eq:PiT_small} and~\eqref{Eq:PiT_large}, we can estimate that the position of the peak of $\mathrm{Re}\Pi_T$ locates at $\omega \sim 2\alpha \mu $. The height of the peak can be estimated as 
$4\mu b_0/(3\pi^2)$, which agrees very well with numerical results. Note that the height of the peak is not suppressed by $v_F/v_s$, and therefore the transverse response can be much larger than the longitudinal one.

Physically, the peak appears because, when the momentum of the axial gauge field is about $\mathbf{p}\approx -2\mathbf{k}_F$, electrons can be back scatted to the same Fermi surface. This Fermi surface nesting effect leads to a large phase space that is responsible for the enhancement of the transverse response function. Due to the spin-momentum locking of Weyl fermions, the longitudinal transition amplitude is vanishingly small if $\mathbf{p}\sim -2\mathbf{k}_F$. Consequently, no peak is observed in the longitudinal response function for the linear dispersion relation case. However, as we will demonstrate subsequently, the longitudinal response also manifests a peak around $p\approx 2k_F$ for external perturbations with quadratic dispersion.

The results for the quadratic dispersion relation are shown in Fig.~\ref{Fig:Pi_d_Re}(c)-(d). Different from the linear dispersion case, the real parts of the response functions in the small frequency limit are now proportional to $\omega$,
\begin{eqnarray}
	&&\mathrm{Re}\Pi_L(\omega,p)\approx \frac{2\gamma b_0 }{\pi^2}\omega, 
\end{eqnarray}
and
\begin{eqnarray}
	&&\mathrm{Re}\Pi_T(\omega,p)\approx \frac{ b_0}{6\pi^2\gamma}\omega.
\end{eqnarray}
Note that with the increasing of frequency, the transverse response increases much faster than the longitudinal one. 
In the large frequency limit, we find that the longitudinal response approaches to a constant
\begin{eqnarray}
&&	\mathrm{Re}\Pi_L(\omega,p)\approx \frac{8\mu b_0\gamma^2}{3\pi^2}+\frac{32\gamma^3\mu^2b_0}{15\pi^2\omega},
\end{eqnarray}
and the transverse one decays as $1/\omega$
\begin{eqnarray} 
&&	\mathrm{Re}\Pi_T(\omega,p)\approx \frac{8\gamma \mu^2 b_0}{3\pi^2 \omega}.
\end{eqnarray}

As can be seen in Fig.~\ref{Fig:Pi_d_Re}(d), for the transverse response there is a peak  located at $\omega\sim 4\gamma \mu$,  and the height of the peak can be estimated as $4\mu b_0/(3\pi^2)$, which is independent of $\gamma$. The longitudinal response also exhibits a peak, see Fig.~\ref{Fig:Pi_d_Re}(c), and the height of the peak is about 
 $4\mu b_0\gamma^2/\pi^2$, which is suppressed by $\gamma^2$.

{\it Hall response.\textemdash}
We now study the 
Hall response $\Pi_H(\omega,p)$.  It is convenient to first compute the real part,
\begin{widetext}
\begin{eqnarray}
	\mathrm{Re}\Pi^{\mathrm{intra}}_{H}(\omega,p)
	&=&\sum_\lambda \lambda \theta(p-|\omega|)\theta(2\mu_\lambda -\omega-p)F_{H}(\omega,\mu_\lambda,p)-(\omega\to -\omega)-(\mu\to -\mu),\\
	\mathrm{Re}\Pi^{\mathrm{inter}}_{H}(\omega,p)
	&=&\sum_\lambda \lambda\theta(|\omega|-p)\theta(p-|\omega+2\mu_\lambda|)F_{H}(-\omega,\mu_\lambda,p)-(\omega\to -\omega),
\end{eqnarray}
\end{widetext}
where $F_{H}(\omega,\mu,p)=(p^2-\omega^2)[p^2-(\omega-2\mu)^2]/(64\pi p^2)$.  The imaginary part is obtained through the Krammers-Kroning relation. Since the real part of the Hall response is odd in frequency, the imaginary part is even.

\begin{figure}
	\includegraphics[width=.43\textwidth]{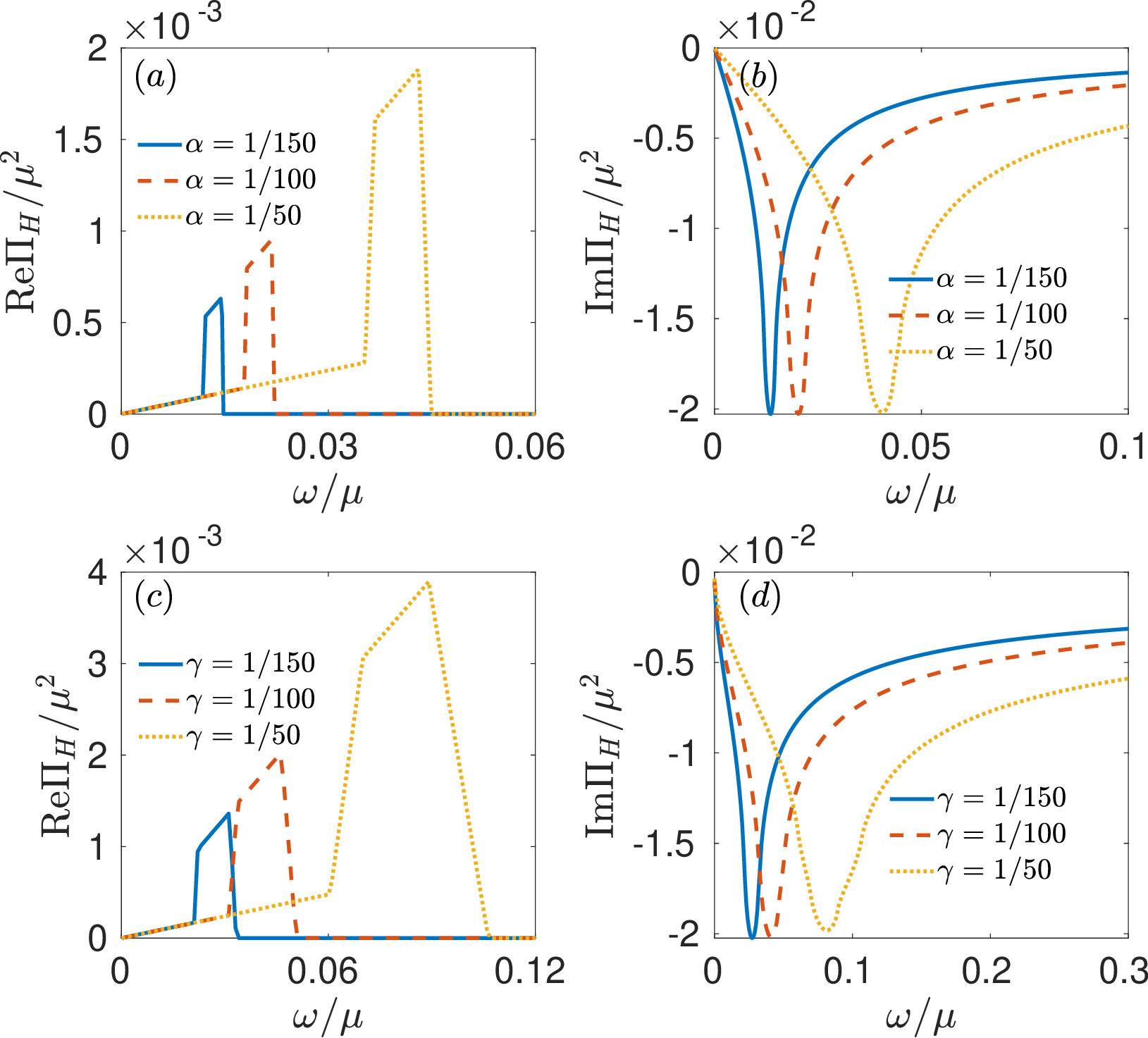}
	\caption{The real and imaginary parts of the off-diagonal response functions for linear [(a)-(b)] and quadratic  [(c)-(d)] dispersion relations. Other parameters are the same as in Fig.~\ref{Fig:Pi_d}. The peaks in (b) and (d) are located at $\omega\sim 2\alpha\mu$ and $\omega\sim 4\gamma\mu$, respectively.}\label{Fig:Pi_H_Re}
\end{figure}

In Fig.~\ref{Fig:Pi_H_Re} we plot the off-diagonal response $\Pi^H_{xy}$ for linear~[(a)-(b)] and quadratic dispersions~[(c)-(d)] as functions of frequency, and one can see that the response functions show similar features for both cases. In the  small frequency limit, the real part increases linearly with frequency and the step function feature is observed with further increasing $\omega$. The real part is typically smaller than the imaginary part, so in the following we focus on the latter. 

For the linear dispersion case and in the  small $\omega$ limit, the imaginary part of the response function increases linearly with $\omega$,
\begin{eqnarray}
	\mathrm{Im}\Pi_{H}(\omega,p)\approx-\frac{b_0\omega}{2\pi^2\alpha},
\end{eqnarray}
while for large $\omega$, it decays as $1/\omega$,
\begin{eqnarray}
	\mathrm{Im}\Pi_{H}(\omega,p)\approx -\frac{2\alpha\mu^2b_0}{\pi^2\omega},
\end{eqnarray}
so the response function reaches a maximal magnitude $2\mu b_0/\pi^2$ at $\omega\approx 2\alpha\mu$, which is similar to the diagonal transverse response function.

For the quadratic dispersion relation, we find that the response function increases as $\sqrt{\omega}$ in the small frequency limit,
\begin{eqnarray}
	\mathrm{Im}\Pi_{H}(\omega,p)\approx-\frac{b_0\sqrt{\mu\omega/\gamma}}{2\pi^2},
\end{eqnarray}
and in the large frequency limit,
\begin{eqnarray}
	\mathrm{Im}\Pi_{H}(\omega,p)\approx-\frac{2b_0\sqrt{\mu^3\gamma}}{\pi^2\sqrt{\omega}},
\end{eqnarray}
which decays as $1/\sqrt{\omega}$. So around $\omega\sim 4\gamma\mu$, it manifests a peak with the magnitude about $2 b_0\mu/\pi^2$.

Similar to the transverse response function, the magnitude of the peak for the Hall response function is determined by the properties of Weyl fermions and is independent of $\alpha$ or $\gamma$.

{\it Effects of disorder.\textemdash}
To explore the effects of disorder, we replace the clean limit Matsubara Green's function by~\cite{altland_simons_2010} $G(i\omega_n,\mathbf{k})=[i\omega_n-H(\mathbf{k})-\mu+i\Gamma\mathrm{sign}(\omega_n)]^{-1}$, where $\Gamma$ is a constant scattering rate. We focus on the transverse and Hall responses, as they are large in the clean limit.

\begin{figure}
	\includegraphics[width=0.43\textwidth]{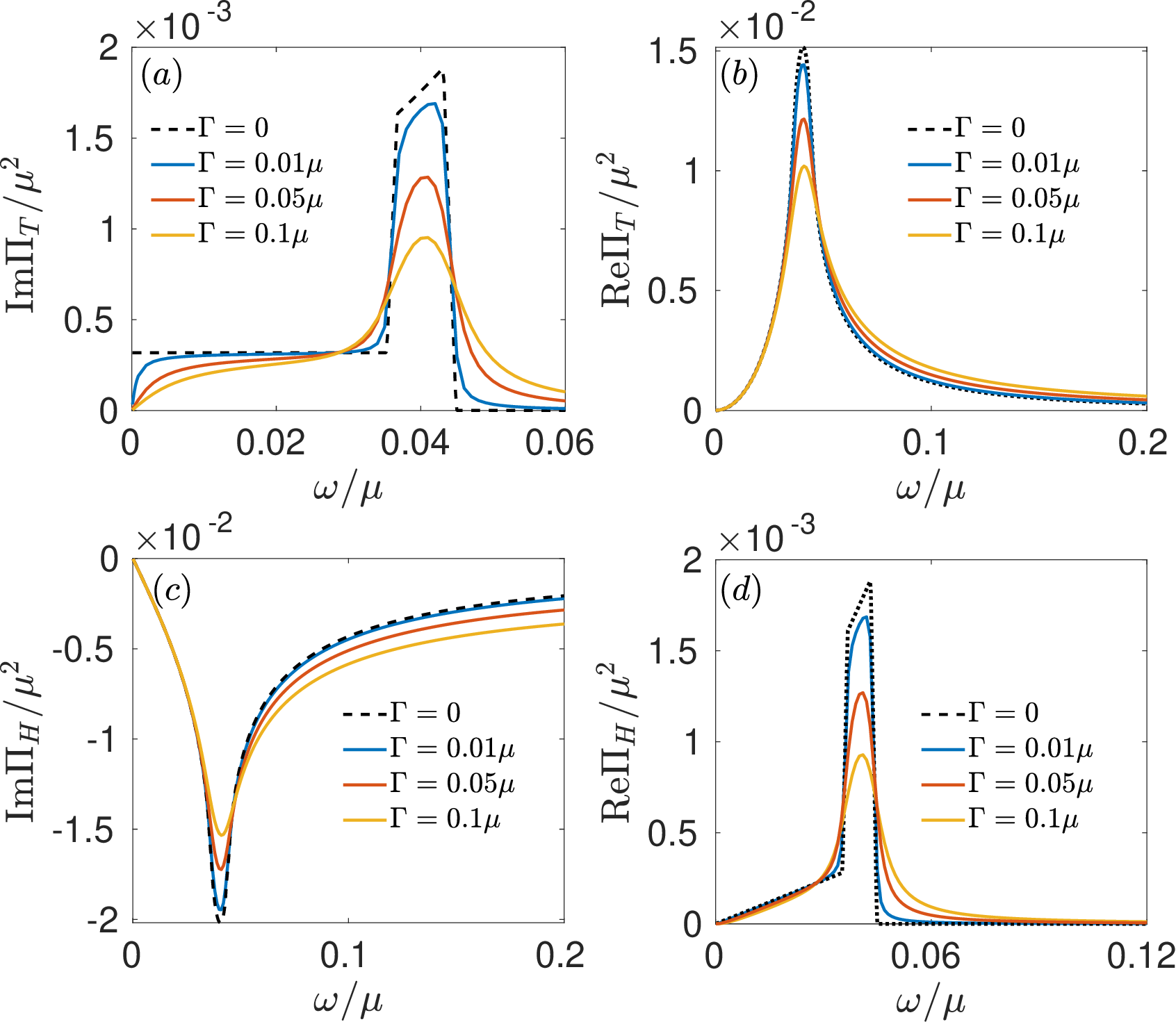}
	\caption{The transverse [(a)-(b)] and Hall [(c)-(d)] response functions in the presence of disorder for the linear dispersion with $\alpha=1/50$ and $b_0=0.1\mu$.}\label{Fig:Dirty}
\end{figure}
Figure~\ref{Fig:Dirty} shows the results for several different disorder strengths $\Gamma$ at fixed $\alpha$. As shown in panels (a) and (d), the step-like feature can be observed for small $\Gamma$ and diminishes gradually as $\Gamma$ increases. However, the peaks are quite robust even for relatively large $\Gamma$, see Fig.~\ref{Fig:Dirty}(b) and (c). Notably, the peak positions are barely changed compared to the clean limit values. We have also performed calculations  for quadratic dispersion and observed qualitatively similar behavior.

{\it Application I: Hall type magnetopiezoelectric effect.\textemdash} Recently, in metals that break both time-reversal and inversion symmetries, a new type of piezoelectricity, the dynamical magnetopiezoelectric effect has been introduced~\cite{PhysRevLett.117.257601,PhysRevB.96.064432,PhysRevB.97.235128} and observed~\cite{PhysRevLett.122.127207}. As an application of our results, we here propose that time-reversal and inversion breaking Weyl semimetals show a Hall type magnetopiezoelectric effect, where an electric current is induced by transverse sound wave, and the direction of the current is perpendicular to the directions of sound propagation and polarization.

Following~\cite{PhysRevLett.115.177202}, strain in Weyl semimetals is related to the  axial gauge field as $A_{5,i}=b u_{iz}+b\delta_{iz}\sum_{j}u_{jj}$, here we have assumed that the Weyl nodes are separated along the $z$-direction, but the expression can be extended to other directions of Weyl nodes separation. The strain tensor is $u_{ij}=(\partial_i u_j +\partial_j u_i)/2$ with $\mathbf{u}(t,\mathbf{r})$ being the displacement field. As an example, we consider a transverse sound wave propagating in the $z$-direction, $\mathbf{u}= u_0\mathbf{e}_x e^{i (q z-\omega t)}$. Then the induced axial gauge field $A_{5,x}\approx iq bu_0e^{i (q z-\omega t)}$ can induce an electric current in the $y$ direction. The Fourier transform of the electric current in the low frequency limit is $j_y=eb_0b q^2u_0/(2\pi^2\hbar)$, where we have recovered $\hbar$ and electric charge $-e$. In noncentrosymmetric magnetic Weyl semimeatals, the Weyl node separation in the momentum space $b$ is of the order of  $0.1\times 2\pi /a$ with $a$ being the lattice constant, and  $b_0$ is of the order of \si{meV}~\cite{PhysRevB.97.041104}. Assuming that the displacement  field $u_0$ is about one percent of the lattice constant and the sound velocity is of the order of $10^5$\si{cm/s}, then the induced current is  estimated as
\begin{eqnarray}
	j_y\approx 0.3\left(\frac{\omega}{2\pi\si{MHz}}\right)^2\frac{\si{\mu A}}{\si{cm^2}}.
\end{eqnarray}
Our order of magnitude estimate indicates that the effect is  large enough to be detected.

{\it Application II: Magnon and electric field interconversion.\textemdash}
The creation and manipulation of magnons are the central focus of magnon spintronics~\cite{Chumak2015}. Our results provide a \revision{possible} mechanism to excite magnons in noncentrosymmetric magnetic Weyls. In these materials, local moments couple to Weyl fermions through $K\mathbf{m}(\mathbf{r})\cdot\mathbf{s}$, where  $K$ is the Kondo coupling constant, $\mathbf{m}$ is the local moments, and $\mathbf{s}$ is the spin operator of Weyl fermions~\cite{Drucker2023,RKKYZhang}. The spin and pseudospin are generally different and their connection is material-dependent. For simplicity, we  assume they are identical~\cite{Drucker2023,RKKYZhang}. Under this approximation, the magnetization $\mathbf{m}$ becomes proportional to the axial gauge potential $\mathbf{A}_5$~\cite{PhysRevB.85.165110,PhysRevB.84.235126} when expanding the Kondo interaction around the Weyl nodes. Thus according to our main results, spin waves can induce electric currents in noncentrosymmetric magnetic Weyl semimetals. This in turn provides a way to excite magnons using electric fields, which is similar to the inverse magnetopiezoelectric effect~\cite{PhysRevB.96.064432}. The transverse response function reaches a maximal around $\omega \sim\alpha \mu$  or $\omega\sim \gamma\mu$. Light with frequencies near $\gamma\mu$ or $\alpha\mu$ could therefore be used to excite magnons; however, further research is required to establish viable experimental setups.

{\it Conclusion.\textemdash}
We investigate the electric current induced by axial gauge fields in time-reversal and inversion breaking Weyl semimetals. We calculate the longitudinal,  transverse, and Hall type response functions for both linear and quadratic dispersion relations of axial gauge potentials.  We find that the transverse and Hall responses are much larger than the longitudinal one. 
With increasing frequency, the transverse and Hall response functions manifest a peak whose height is determined by the parameters of Weyl fermions and is independent of external perturbations. We additionally demonstrate that the essential features of the response functions survive in the presence of disorder.
As applications of our results, we predict a Hall type magnetopiezoelectric effect, where a sizable electric current can be generated by dynamical strain. Additionally, we propose a possible  mechanism to excite magnons using electric fields in  noncentrosymmetric magnetic Weyl semimetals and this could be exploited for magnon spintronics.

\emph{Acknowledgments.\textemdash}We thank P. O. Sukhachov and Xi Luo for useful discussions. This work is supported by National Natural Science
Foundation of China (Grant No.~12204329).

\bibliography{ACME.bib}

\begin{thebibliography}{56}%
\makeatletter
\providecommand \@ifxundefined [1]{%
 \@ifx{#1\undefined}
}%
\providecommand \@ifnum [1]{%
 \ifnum #1\expandafter \@firstoftwo
 \else \expandafter \@secondoftwo
 \fi
}%
\providecommand \@ifx [1]{%
 \ifx #1\expandafter \@firstoftwo
 \else \expandafter \@secondoftwo
 \fi
}%
\providecommand \natexlab [1]{#1}%
\providecommand \enquote  [1]{``#1''}%
\providecommand \bibnamefont  [1]{#1}%
\providecommand \bibfnamefont [1]{#1}%
\providecommand \citenamefont [1]{#1}%
\providecommand \href@noop [0]{\@secondoftwo}%
\providecommand \href [0]{\begingroup \@sanitize@url \@href}%
\providecommand \@href[1]{\@@startlink{#1}\@@href}%
\providecommand \@@href[1]{\endgroup#1\@@endlink}%
\providecommand \@sanitize@url [0]{\catcode `\\12\catcode `\$12\catcode
  `\&12\catcode `\#12\catcode `\^12\catcode `\_12\catcode `\%12\relax}%
\providecommand \@@startlink[1]{}%
\providecommand \@@endlink[0]{}%
\providecommand \url  [0]{\begingroup\@sanitize@url \@url }%
\providecommand \@url [1]{\endgroup\@href {#1}{\urlprefix }}%
\providecommand \urlprefix  [0]{URL }%
\providecommand \Eprint [0]{\href }%
\providecommand \doibase [0]{https://doi.org/}%
\providecommand \selectlanguage [0]{\@gobble}%
\providecommand \bibinfo  [0]{\@secondoftwo}%
\providecommand \bibfield  [0]{\@secondoftwo}%
\providecommand \translation [1]{[#1]}%
\providecommand \BibitemOpen [0]{}%
\providecommand \bibitemStop [0]{}%
\providecommand \bibitemNoStop [0]{.\EOS\space}%
\providecommand \EOS [0]{\spacefactor3000\relax}%
\providecommand \BibitemShut  [1]{\csname bibitem#1\endcsname}%
\let\auto@bib@innerbib\@empty
\bibitem [{\citenamefont {Armitage}\ \emph {et~al.}(2018)\citenamefont
  {Armitage}, \citenamefont {Mele},\ and\ \citenamefont
  {Vishwanath}}]{Armitage:rev-2018}%
  \BibitemOpen
  \bibfield  {author} {\bibinfo {author} {\bibfnamefont {N.~P.}\ \bibnamefont
  {Armitage}}, \bibinfo {author} {\bibfnamefont {E.~J.}\ \bibnamefont {Mele}},\
  and\ \bibinfo {author} {\bibfnamefont {A.}~\bibnamefont {Vishwanath}},\
  }\bibfield  {title} {\bibinfo {title} {{W}eyl and {D}irac semimetals in
  three-dimensional solids},\ }\href
  {https://doi.org/10.1103/RevModPhys.90.015001} {\bibfield  {journal}
  {\bibinfo  {journal} {Rev. Mod. Phys.}\ }\textbf {\bibinfo {volume} {90}},\
  \bibinfo {pages} {015001} (\bibinfo {year} {2018})}\BibitemShut {NoStop}%
\bibitem [{\citenamefont {Lv}\ \emph {et~al.}(2021)\citenamefont {Lv},
  \citenamefont {Qian},\ and\ \citenamefont {Ding}}]{HongDingRMP}%
  \BibitemOpen
  \bibfield  {author} {\bibinfo {author} {\bibfnamefont {B.~Q.}\ \bibnamefont
  {Lv}}, \bibinfo {author} {\bibfnamefont {T.}~\bibnamefont {Qian}},\ and\
  \bibinfo {author} {\bibfnamefont {H.}~\bibnamefont {Ding}},\ }\bibfield
  {title} {\bibinfo {title} {Experimental perspective on three-dimensional
  topological semimetals},\ }\href
  {https://doi.org/10.1103/RevModPhys.93.025002} {\bibfield  {journal}
  {\bibinfo  {journal} {Rev. Mod. Phys.}\ }\textbf {\bibinfo {volume} {93}},\
  \bibinfo {pages} {025002} (\bibinfo {year} {2021})}\BibitemShut {NoStop}%
\bibitem [{\citenamefont {Ilan}\ \emph {et~al.}(2020)\citenamefont {Ilan},
  \citenamefont {Grushin},\ and\ \citenamefont
  {Pikulin}}]{Ilan-Pikulin:rev-2019}%
  \BibitemOpen
  \bibfield  {author} {\bibinfo {author} {\bibfnamefont {R.}~\bibnamefont
  {Ilan}}, \bibinfo {author} {\bibfnamefont {A.~G.}\ \bibnamefont {Grushin}},\
  and\ \bibinfo {author} {\bibfnamefont {D.~I.}\ \bibnamefont {Pikulin}},\
  }\bibfield  {title} {\bibinfo {title} {Pseudo-electromagnetic fields in {3D}
  topological semimetals},\ }\href {https://doi.org/10.1038/s42254-019-0121-8}
  {\bibfield  {journal} {\bibinfo  {journal} {Nat. Rev. Phys.}\ }\textbf
  {\bibinfo {volume} {2}},\ \bibinfo {pages} {29} (\bibinfo {year}
  {2020})}\BibitemShut {NoStop}%
\bibitem [{\citenamefont {Yu}\ and\ \citenamefont {Liu}(2021)}]{Yu_2021}%
  \BibitemOpen
  \bibfield  {author} {\bibinfo {author} {\bibfnamefont {J.}~\bibnamefont
  {Yu}}\ and\ \bibinfo {author} {\bibfnamefont {C.-X.}\ \bibnamefont {Liu}},\
  }\bibinfo {title} {Pseudo-gauge fields in {Dirac and Weyl} materials},\ in\
  \href {https://doi.org/10.1016/bs.semsem.2021.06.003} {\emph {\bibinfo
  {booktitle} {Topological Insulator and Related Topics}}}\ (\bibinfo
  {publisher} {Elsevier},\ \bibinfo {year} {2021})\ p.\ \bibinfo {pages}
  {195–224}\BibitemShut {NoStop}%
\bibitem [{\citenamefont {Jadidi}\ \emph {et~al.}(2020)\citenamefont {Jadidi},
  \citenamefont {Kargarian}, \citenamefont {Mittendorff}, \citenamefont
  {Aytac}, \citenamefont {Shen}, \citenamefont {K\"onig-Otto}, \citenamefont
  {Winnerl}, \citenamefont {Ni}, \citenamefont {Gaeta}, \citenamefont
  {Murphy},\ and\ \citenamefont {Drew}}]{PhysRevB.102.245123}%
  \BibitemOpen
  \bibfield  {author} {\bibinfo {author} {\bibfnamefont {M.~M.}\ \bibnamefont
  {Jadidi}}, \bibinfo {author} {\bibfnamefont {M.}~\bibnamefont {Kargarian}},
  \bibinfo {author} {\bibfnamefont {M.}~\bibnamefont {Mittendorff}}, \bibinfo
  {author} {\bibfnamefont {Y.}~\bibnamefont {Aytac}}, \bibinfo {author}
  {\bibfnamefont {B.}~\bibnamefont {Shen}}, \bibinfo {author} {\bibfnamefont
  {J.~C.}\ \bibnamefont {K\"onig-Otto}}, \bibinfo {author} {\bibfnamefont
  {S.}~\bibnamefont {Winnerl}}, \bibinfo {author} {\bibfnamefont
  {N.}~\bibnamefont {Ni}}, \bibinfo {author} {\bibfnamefont {A.~L.}\
  \bibnamefont {Gaeta}}, \bibinfo {author} {\bibfnamefont {T.~E.}\ \bibnamefont
  {Murphy}},\ and\ \bibinfo {author} {\bibfnamefont {H.~D.}\ \bibnamefont
  {Drew}},\ }\bibfield  {title} {\bibinfo {title} {Nonlinear optical control of
  chiral charge pumping in a topological {W}eyl semimetal},\ }\href
  {https://doi.org/10.1103/PhysRevB.102.245123} {\bibfield  {journal} {\bibinfo
   {journal} {Phys. Rev. B}\ }\textbf {\bibinfo {volume} {102}},\ \bibinfo
  {pages} {245123} (\bibinfo {year} {2020})}\BibitemShut {NoStop}%
\bibitem [{\citenamefont {Nandy}\ and\ \citenamefont
  {Pesin}(2020)}]{PhysRevLett.125.266601}%
  \BibitemOpen
  \bibfield  {author} {\bibinfo {author} {\bibfnamefont {S.}~\bibnamefont
  {Nandy}}\ and\ \bibinfo {author} {\bibfnamefont {D.~A.}\ \bibnamefont
  {Pesin}},\ }\bibfield  {title} {\bibinfo {title} {Chiral magnetic effect of
  hot electrons},\ }\href {https://doi.org/10.1103/PhysRevLett.125.266601}
  {\bibfield  {journal} {\bibinfo  {journal} {Phys. Rev. Lett.}\ }\textbf
  {\bibinfo {volume} {125}},\ \bibinfo {pages} {266601} (\bibinfo {year}
  {2020})}\BibitemShut {NoStop}%
\bibitem [{\citenamefont {Song}\ \emph {et~al.}(2016)\citenamefont {Song},
  \citenamefont {Zhao}, \citenamefont {Fang},\ and\ \citenamefont
  {Dai}}]{PhysRevB.94.214306}%
  \BibitemOpen
  \bibfield  {author} {\bibinfo {author} {\bibfnamefont {Z.}~\bibnamefont
  {Song}}, \bibinfo {author} {\bibfnamefont {J.}~\bibnamefont {Zhao}}, \bibinfo
  {author} {\bibfnamefont {Z.}~\bibnamefont {Fang}},\ and\ \bibinfo {author}
  {\bibfnamefont {X.}~\bibnamefont {Dai}},\ }\bibfield  {title} {\bibinfo
  {title} {Detecting the chiral magnetic effect by lattice dynamics in {W}eyl
  semimetals},\ }\href {https://doi.org/10.1103/PhysRevB.94.214306} {\bibfield
  {journal} {\bibinfo  {journal} {Phys. Rev. B}\ }\textbf {\bibinfo {volume}
  {94}},\ \bibinfo {pages} {214306} (\bibinfo {year} {2016})}\BibitemShut
  {NoStop}%
\bibitem [{\citenamefont {Yuan}\ \emph {et~al.}(2020)\citenamefont {Yuan},
  \citenamefont {Zhang}, \citenamefont {Zhang}, \citenamefont {Yan},
  \citenamefont {Lyu}, \citenamefont {Zhang}, \citenamefont {Li}, \citenamefont
  {Song}, \citenamefont {Zhao}, \citenamefont {Leng}, \citenamefont {Ozerov},
  \citenamefont {Chen}, \citenamefont {Wang}, \citenamefont {Shi},
  \citenamefont {Yan},\ and\ \citenamefont {Xiu}}]{Yuan2020}%
  \BibitemOpen
  \bibfield  {author} {\bibinfo {author} {\bibfnamefont {X.}~\bibnamefont
  {Yuan}}, \bibinfo {author} {\bibfnamefont {C.}~\bibnamefont {Zhang}},
  \bibinfo {author} {\bibfnamefont {Y.}~\bibnamefont {Zhang}}, \bibinfo
  {author} {\bibfnamefont {Z.}~\bibnamefont {Yan}}, \bibinfo {author}
  {\bibfnamefont {T.}~\bibnamefont {Lyu}}, \bibinfo {author} {\bibfnamefont
  {M.}~\bibnamefont {Zhang}}, \bibinfo {author} {\bibfnamefont
  {Z.}~\bibnamefont {Li}}, \bibinfo {author} {\bibfnamefont {C.}~\bibnamefont
  {Song}}, \bibinfo {author} {\bibfnamefont {M.}~\bibnamefont {Zhao}}, \bibinfo
  {author} {\bibfnamefont {P.}~\bibnamefont {Leng}}, \bibinfo {author}
  {\bibfnamefont {M.}~\bibnamefont {Ozerov}}, \bibinfo {author} {\bibfnamefont
  {X.}~\bibnamefont {Chen}}, \bibinfo {author} {\bibfnamefont {N.}~\bibnamefont
  {Wang}}, \bibinfo {author} {\bibfnamefont {Y.}~\bibnamefont {Shi}}, \bibinfo
  {author} {\bibfnamefont {H.}~\bibnamefont {Yan}},\ and\ \bibinfo {author}
  {\bibfnamefont {F.}~\bibnamefont {Xiu}},\ }\bibfield  {title} {\bibinfo
  {title} {The discovery of dynamic chiral anomaly in a {W}eyl semimetal
  {NbAs}},\ }\href {https://doi.org/10.1038/s41467-020-14749-4} {\bibfield
  {journal} {\bibinfo  {journal} {Nat. Commun.}\ }\textbf {\bibinfo {volume}
  {11}},\ \bibinfo {pages} {1259} (\bibinfo {year} {2020})}\BibitemShut
  {NoStop}%
\bibitem [{\citenamefont {Cortijo}\ \emph
  {et~al.}(2016{\natexlab{a}})\citenamefont {Cortijo}, \citenamefont
  {Kharzeev}, \citenamefont {Landsteiner},\ and\ \citenamefont
  {Vozmediano}}]{PhysRevB.94.241405}%
  \BibitemOpen
  \bibfield  {author} {\bibinfo {author} {\bibfnamefont {A.}~\bibnamefont
  {Cortijo}}, \bibinfo {author} {\bibfnamefont {D.}~\bibnamefont {Kharzeev}},
  \bibinfo {author} {\bibfnamefont {K.}~\bibnamefont {Landsteiner}},\ and\
  \bibinfo {author} {\bibfnamefont {M.~A.~H.}\ \bibnamefont {Vozmediano}},\
  }\bibfield  {title} {\bibinfo {title} {Strain-induced chiral magnetic effect
  in {W}eyl semimetals},\ }\href {https://doi.org/10.1103/PhysRevB.94.241405}
  {\bibfield  {journal} {\bibinfo  {journal} {Phys. Rev. B}\ }\textbf {\bibinfo
  {volume} {94}},\ \bibinfo {pages} {241405} (\bibinfo {year}
  {2016}{\natexlab{a}})}\BibitemShut {NoStop}%
\bibitem [{\citenamefont {Diaz}\ \emph {et~al.}(2021)\citenamefont {Diaz},
  \citenamefont {Putzke}, \citenamefont {Huang}, \citenamefont {Estry},
  \citenamefont {Analytis}, \citenamefont {Sabsovich}, \citenamefont {Grushin},
  \citenamefont {Ilan},\ and\ \citenamefont {Moll}}]{pub.1142612917}%
  \BibitemOpen
  \bibfield  {author} {\bibinfo {author} {\bibfnamefont {J.}~\bibnamefont
  {Diaz}}, \bibinfo {author} {\bibfnamefont {C.}~\bibnamefont {Putzke}},
  \bibinfo {author} {\bibfnamefont {X.}~\bibnamefont {Huang}}, \bibinfo
  {author} {\bibfnamefont {A.}~\bibnamefont {Estry}}, \bibinfo {author}
  {\bibfnamefont {J.~G.}\ \bibnamefont {Analytis}}, \bibinfo {author}
  {\bibfnamefont {D.}~\bibnamefont {Sabsovich}}, \bibinfo {author}
  {\bibfnamefont {A.~G.}\ \bibnamefont {Grushin}}, \bibinfo {author}
  {\bibfnamefont {R.}~\bibnamefont {Ilan}},\ and\ \bibinfo {author}
  {\bibfnamefont {P.~J.~W.}\ \bibnamefont {Moll}},\ }\bibfield  {title}
  {\bibinfo {title} {Bending strain in 3{D} topological semi-metals},\ }\href
  {https://doi.org/10.1088/1361-6463/ac357f} {\bibfield  {journal} {\bibinfo
  {journal} {J. Phys. D: Appl. Phys.}\ }\textbf {\bibinfo {volume} {55}},\
  \bibinfo {pages} {084001} (\bibinfo {year} {2021})}\BibitemShut {NoStop}%
\bibitem [{\citenamefont {Zyuzin}\ \emph {et~al.}(2012)\citenamefont {Zyuzin},
  \citenamefont {Wu},\ and\ \citenamefont {Burkov}}]{PhysRevB.85.165110}%
  \BibitemOpen
  \bibfield  {author} {\bibinfo {author} {\bibfnamefont {A.~A.}\ \bibnamefont
  {Zyuzin}}, \bibinfo {author} {\bibfnamefont {S.}~\bibnamefont {Wu}},\ and\
  \bibinfo {author} {\bibfnamefont {A.~A.}\ \bibnamefont {Burkov}},\ }\bibfield
   {title} {\bibinfo {title} {Weyl semimetal with broken time reversal and
  inversion symmetries},\ }\href {https://doi.org/10.1103/PhysRevB.85.165110}
  {\bibfield  {journal} {\bibinfo  {journal} {Phys. Rev. B}\ }\textbf {\bibinfo
  {volume} {85}},\ \bibinfo {pages} {165110} (\bibinfo {year}
  {2012})}\BibitemShut {NoStop}%
\bibitem [{\citenamefont {Son}\ and\ \citenamefont
  {Yamamoto}(2012)}]{PhysRevLett.109.181602}%
  \BibitemOpen
  \bibfield  {author} {\bibinfo {author} {\bibfnamefont {D.~T.}\ \bibnamefont
  {Son}}\ and\ \bibinfo {author} {\bibfnamefont {N.}~\bibnamefont {Yamamoto}},\
  }\bibfield  {title} {\bibinfo {title} {Berry curvature, triangle anomalies,
  and the chiral magnetic effect in {F}ermi liquids},\ }\href
  {https://doi.org/10.1103/PhysRevLett.109.181602} {\bibfield  {journal}
  {\bibinfo  {journal} {Phys. Rev. Lett.}\ }\textbf {\bibinfo {volume} {109}},\
  \bibinfo {pages} {181602} (\bibinfo {year} {2012})}\BibitemShut {NoStop}%
\bibitem [{\citenamefont {Zhou}\ \emph {et~al.}(2013)\citenamefont {Zhou},
  \citenamefont {Jiang}, \citenamefont {Niu},\ and\ \citenamefont
  {Shi}}]{Zhou_2013}%
  \BibitemOpen
  \bibfield  {author} {\bibinfo {author} {\bibfnamefont {J.-H.}\ \bibnamefont
  {Zhou}}, \bibinfo {author} {\bibfnamefont {H.}~\bibnamefont {Jiang}},
  \bibinfo {author} {\bibfnamefont {Q.}~\bibnamefont {Niu}},\ and\ \bibinfo
  {author} {\bibfnamefont {J.-R.}\ \bibnamefont {Shi}},\ }\bibfield  {title}
  {\bibinfo {title} {Topological invariants of metals and the related physical
  effects},\ }\href {https://doi.org/10.1088/0256-307x/30/2/027101} {\bibfield
  {journal} {\bibinfo  {journal} {Chin. Phys. Lett}\ }\textbf {\bibinfo
  {volume} {30}},\ \bibinfo {pages} {027101} (\bibinfo {year}
  {2013})}\BibitemShut {NoStop}%
\bibitem [{\citenamefont {Ashby}\ and\ \citenamefont
  {Carbotte}(2013)}]{PhysRevB.87.245131}%
  \BibitemOpen
  \bibfield  {author} {\bibinfo {author} {\bibfnamefont {P.~E.~C.}\
  \bibnamefont {Ashby}}\ and\ \bibinfo {author} {\bibfnamefont {J.~P.}\
  \bibnamefont {Carbotte}},\ }\bibfield  {title} {\bibinfo {title}
  {Magneto-optical conductivity of {W}eyl semimetals},\ }\href
  {https://doi.org/10.1103/PhysRevB.87.245131} {\bibfield  {journal} {\bibinfo
  {journal} {Phys. Rev. B}\ }\textbf {\bibinfo {volume} {87}},\ \bibinfo
  {pages} {245131} (\bibinfo {year} {2013})}\BibitemShut {NoStop}%
\bibitem [{\citenamefont {Tabert}\ and\ \citenamefont
  {Carbotte}(2016)}]{PhysRevB.93.085442}%
  \BibitemOpen
  \bibfield  {author} {\bibinfo {author} {\bibfnamefont {C.~J.}\ \bibnamefont
  {Tabert}}\ and\ \bibinfo {author} {\bibfnamefont {J.~P.}\ \bibnamefont
  {Carbotte}},\ }\bibfield  {title} {\bibinfo {title} {Optical conductivity of
  {W}eyl semimetals and signatures of the gapped semimetal phase transition},\
  }\href {https://doi.org/10.1103/PhysRevB.93.085442} {\bibfield  {journal}
  {\bibinfo  {journal} {Phys. Rev. B}\ }\textbf {\bibinfo {volume} {93}},\
  \bibinfo {pages} {085442} (\bibinfo {year} {2016})}\BibitemShut {NoStop}%
\bibitem [{\citenamefont {Carbotte}(2016)}]{PhysRevB.94.165111}%
  \BibitemOpen
  \bibfield  {author} {\bibinfo {author} {\bibfnamefont {J.~P.}\ \bibnamefont
  {Carbotte}},\ }\bibfield  {title} {\bibinfo {title} {{Dirac cone tilt on
  interband optical background of type-I and type-II Weyl semimetals}},\ }\href
  {https://doi.org/10.1103/PhysRevB.94.165111} {\bibfield  {journal} {\bibinfo
  {journal} {Phys. Rev. B}\ }\textbf {\bibinfo {volume} {94}},\ \bibinfo
  {pages} {165111} (\bibinfo {year} {2016})}\BibitemShut {NoStop}%
\bibitem [{\citenamefont {Roy}\ and\ \citenamefont {Juri\ifmmode \check{c}\else
  \v{c}\fi{}i\ifmmode~\acute{c}\else \'{c}\fi{}}(2017)}]{PhysRevB.96.155117}%
  \BibitemOpen
  \bibfield  {author} {\bibinfo {author} {\bibfnamefont {B.}~\bibnamefont
  {Roy}}\ and\ \bibinfo {author} {\bibfnamefont {V.}~\bibnamefont {Juri\ifmmode
  \check{c}\else \v{c}\fi{}i\ifmmode~\acute{c}\else \'{c}\fi{}}},\ }\bibfield
  {title} {\bibinfo {title} {Optical conductivity of an interacting {W}eyl
  liquid in the collisionless regime},\ }\href
  {https://doi.org/10.1103/PhysRevB.96.155117} {\bibfield  {journal} {\bibinfo
  {journal} {Phys. Rev. B}\ }\textbf {\bibinfo {volume} {96}},\ \bibinfo
  {pages} {155117} (\bibinfo {year} {2017})}\BibitemShut {NoStop}%
\bibitem [{\citenamefont {St\aa{}lhammar}\ \emph {et~al.}(2020)\citenamefont
  {St\aa{}lhammar}, \citenamefont {Larana-Aragon}, \citenamefont {Knolle},\
  and\ \citenamefont {Bergholtz}}]{PhysRevB.102.235134}%
  \BibitemOpen
  \bibfield  {author} {\bibinfo {author} {\bibfnamefont {M.}~\bibnamefont
  {St\aa{}lhammar}}, \bibinfo {author} {\bibfnamefont {J.}~\bibnamefont
  {Larana-Aragon}}, \bibinfo {author} {\bibfnamefont {J.}~\bibnamefont
  {Knolle}},\ and\ \bibinfo {author} {\bibfnamefont {E.~J.}\ \bibnamefont
  {Bergholtz}},\ }\bibfield  {title} {\bibinfo {title} {Magneto-optical
  conductivity in generic {W}eyl semimetals},\ }\href
  {https://doi.org/10.1103/PhysRevB.102.235134} {\bibfield  {journal} {\bibinfo
   {journal} {Phys. Rev. B}\ }\textbf {\bibinfo {volume} {102}},\ \bibinfo
  {pages} {235134} (\bibinfo {year} {2020})}\BibitemShut {NoStop}%
\bibitem [{\citenamefont {Acheche}\ \emph {et~al.}(2020)\citenamefont
  {Acheche}, \citenamefont {Nourafkan}, \citenamefont {Padayasi}, \citenamefont
  {Martin},\ and\ \citenamefont {Tremblay}}]{PhysRevB.102.045148}%
  \BibitemOpen
  \bibfield  {author} {\bibinfo {author} {\bibfnamefont {S.}~\bibnamefont
  {Acheche}}, \bibinfo {author} {\bibfnamefont {R.}~\bibnamefont {Nourafkan}},
  \bibinfo {author} {\bibfnamefont {J.}~\bibnamefont {Padayasi}}, \bibinfo
  {author} {\bibfnamefont {N.}~\bibnamefont {Martin}},\ and\ \bibinfo {author}
  {\bibfnamefont {A.-M.~S.}\ \bibnamefont {Tremblay}},\ }\bibfield  {title}
  {\bibinfo {title} {Interaction and temperature effects on the magneto-optical
  conductivity of {W}eyl liquids},\ }\href
  {https://doi.org/10.1103/PhysRevB.102.045148} {\bibfield  {journal} {\bibinfo
   {journal} {Phys. Rev. B}\ }\textbf {\bibinfo {volume} {102}},\ \bibinfo
  {pages} {045148} (\bibinfo {year} {2020})}\BibitemShut {NoStop}%
\bibitem [{\citenamefont {Hou}\ \emph {et~al.}(2023)\citenamefont {Hou},
  \citenamefont {Yan}, \citenamefont {Tan}, \citenamefont {Li}, \citenamefont
  {Wang}, \citenamefont {Guo},\ and\ \citenamefont
  {Chang}}]{PhysRevB.108.035407}%
  \BibitemOpen
  \bibfield  {author} {\bibinfo {author} {\bibfnamefont {J.-T.}\ \bibnamefont
  {Hou}}, \bibinfo {author} {\bibfnamefont {C.-X.}\ \bibnamefont {Yan}},
  \bibinfo {author} {\bibfnamefont {C.-Y.}\ \bibnamefont {Tan}}, \bibinfo
  {author} {\bibfnamefont {Z.-Q.}\ \bibnamefont {Li}}, \bibinfo {author}
  {\bibfnamefont {P.}~\bibnamefont {Wang}}, \bibinfo {author} {\bibfnamefont
  {H.}~\bibnamefont {Guo}},\ and\ \bibinfo {author} {\bibfnamefont {H.-R.}\
  \bibnamefont {Chang}},\ }\bibfield  {title} {\bibinfo {title} {Effects of
  spatial dimensionality and band tilting on the longitudinal optical
  conductivities in {D}irac bands},\ }\href
  {https://doi.org/10.1103/PhysRevB.108.035407} {\bibfield  {journal} {\bibinfo
   {journal} {Phys. Rev. B}\ }\textbf {\bibinfo {volume} {108}},\ \bibinfo
  {pages} {035407} (\bibinfo {year} {2023})}\BibitemShut {NoStop}%
\bibitem [{\citenamefont {Mukherjee}\ and\ \citenamefont
  {Carbotte}(2018)}]{PhysRevB.97.035144}%
  \BibitemOpen
  \bibfield  {author} {\bibinfo {author} {\bibfnamefont {S.~P.}\ \bibnamefont
  {Mukherjee}}\ and\ \bibinfo {author} {\bibfnamefont {J.~P.}\ \bibnamefont
  {Carbotte}},\ }\bibfield  {title} {\bibinfo {title} {Imaginary part of {H}all
  conductivity in a tilted doped weyl semimetal with both broken time-reversal
  and inversion symmetry},\ }\href {https://doi.org/10.1103/PhysRevB.97.035144}
  {\bibfield  {journal} {\bibinfo  {journal} {Phys. Rev. B}\ }\textbf {\bibinfo
  {volume} {97}},\ \bibinfo {pages} {035144} (\bibinfo {year}
  {2018})}\BibitemShut {NoStop}%
\bibitem [{\citenamefont {Cortijo}\ \emph
  {et~al.}(2016{\natexlab{b}})\citenamefont {Cortijo}, \citenamefont
  {Kharzeev}, \citenamefont {Landsteiner},\ and\ \citenamefont
  {Vozmediano}}]{Cortijo-Vozmediano:2016}%
  \BibitemOpen
  \bibfield  {author} {\bibinfo {author} {\bibfnamefont {A.}~\bibnamefont
  {Cortijo}}, \bibinfo {author} {\bibfnamefont {D.}~\bibnamefont {Kharzeev}},
  \bibinfo {author} {\bibfnamefont {K.}~\bibnamefont {Landsteiner}},\ and\
  \bibinfo {author} {\bibfnamefont {M.~A.~H.}\ \bibnamefont {Vozmediano}},\
  }\bibfield  {title} {\bibinfo {title} {Strain-induced chiral magnetic effect
  in {W}eyl semimetals},\ }\href {https://doi.org/10.1103/PhysRevB.94.241405}
  {\bibfield  {journal} {\bibinfo  {journal} {Phys. Rev. B}\ }\textbf {\bibinfo
  {volume} {94}},\ \bibinfo {pages} {241405} (\bibinfo {year}
  {2016}{\natexlab{b}})}\BibitemShut {NoStop}%
\bibitem [{\citenamefont {Pikulin}\ \emph {et~al.}(2016)\citenamefont
  {Pikulin}, \citenamefont {Chen},\ and\ \citenamefont
  {Franz}}]{Pikulin-Franz:2016}%
  \BibitemOpen
  \bibfield  {author} {\bibinfo {author} {\bibfnamefont {D.~I.}\ \bibnamefont
  {Pikulin}}, \bibinfo {author} {\bibfnamefont {A.}~\bibnamefont {Chen}},\ and\
  \bibinfo {author} {\bibfnamefont {M.}~\bibnamefont {Franz}},\ }\bibfield
  {title} {\bibinfo {title} {Chiral anomaly from strain-induced gauge fields in
  {D}irac and {W}eyl semimetals},\ }\href
  {https://doi.org/10.1103/PhysRevX.6.041021} {\bibfield  {journal} {\bibinfo
  {journal} {Phys. Rev. X}\ }\textbf {\bibinfo {volume} {6}},\ \bibinfo {pages}
  {041021} (\bibinfo {year} {2016})}\BibitemShut {NoStop}%
\bibitem [{\citenamefont {Grushin}\ \emph {et~al.}(2016)\citenamefont
  {Grushin}, \citenamefont {Venderbos}, \citenamefont {Vishwanath},\ and\
  \citenamefont {Ilan}}]{Grushin:2016}%
  \BibitemOpen
  \bibfield  {author} {\bibinfo {author} {\bibfnamefont {A.~G.}\ \bibnamefont
  {Grushin}}, \bibinfo {author} {\bibfnamefont {J.~W.}\ \bibnamefont
  {Venderbos}}, \bibinfo {author} {\bibfnamefont {A.}~\bibnamefont
  {Vishwanath}},\ and\ \bibinfo {author} {\bibfnamefont {R.}~\bibnamefont
  {Ilan}},\ }\bibfield  {title} {\bibinfo {title} {Inhomogeneous {W}eyl and
  {D}irac semimetals: Transport in axial magnetic fields and {F}ermi arc
  surface states from pseudo-{L}andau levels},\ }\href
  {https://doi.org/10.1103/PhysRevX.6.041046} {\bibfield  {journal} {\bibinfo
  {journal} {Phys. Rev. X}\ }\textbf {\bibinfo {volume} {6}},\ \bibinfo {pages}
  {041046} (\bibinfo {year} {2016})}\BibitemShut {NoStop}%
\bibitem [{\citenamefont {Landsteiner}(2016)}]{Landsteiner_2016}%
  \BibitemOpen
  \bibfield  {author} {\bibinfo {author} {\bibfnamefont {K.}~\bibnamefont
  {Landsteiner}},\ }\bibfield  {title} {\bibinfo {title} {Notes on anomaly
  induced transport},\ }\href {http://dx.doi.org/10.5506/APhysPolB.47.2617}
  {\bibfield  {journal} {\bibinfo  {journal} {Acta Phys. Pol. B}\ }\textbf
  {\bibinfo {volume} {47}},\ \bibinfo {pages} {2617} (\bibinfo {year}
  {2016})}\BibitemShut {NoStop}%
\bibitem [{\citenamefont {Huang}\ \emph {et~al.}(2017)\citenamefont {Huang},
  \citenamefont {Zhou},\ and\ \citenamefont {Shen}}]{PhysRevB.96.085201}%
  \BibitemOpen
  \bibfield  {author} {\bibinfo {author} {\bibfnamefont {Z.-M.}\ \bibnamefont
  {Huang}}, \bibinfo {author} {\bibfnamefont {J.}~\bibnamefont {Zhou}},\ and\
  \bibinfo {author} {\bibfnamefont {S.-Q.}\ \bibnamefont {Shen}},\ }\bibfield
  {title} {\bibinfo {title} {Topological responses from chiral anomaly in
  multi-{W}eyl semimetals},\ }\href
  {https://doi.org/10.1103/PhysRevB.96.085201} {\bibfield  {journal} {\bibinfo
  {journal} {Phys. Rev. B}\ }\textbf {\bibinfo {volume} {96}},\ \bibinfo
  {pages} {085201} (\bibinfo {year} {2017})}\BibitemShut {NoStop}%
\bibitem [{\citenamefont {Weststr\"om}\ and\ \citenamefont
  {Ojanen}(2017)}]{PhysRevX.7.041026}%
  \BibitemOpen
  \bibfield  {author} {\bibinfo {author} {\bibfnamefont {A.}~\bibnamefont
  {Weststr\"om}}\ and\ \bibinfo {author} {\bibfnamefont {T.}~\bibnamefont
  {Ojanen}},\ }\bibfield  {title} {\bibinfo {title} {Designer curved-space
  geometry for relativistic fermions in {W}eyl metamaterials},\ }\href
  {https://doi.org/10.1103/PhysRevX.7.041026} {\bibfield  {journal} {\bibinfo
  {journal} {Phys. Rev. X}\ }\textbf {\bibinfo {volume} {7}},\ \bibinfo {pages}
  {041026} (\bibinfo {year} {2017})}\BibitemShut {NoStop}%
\bibitem [{\citenamefont {Chernodub}\ and\ \citenamefont
  {Zubkov}(2017)}]{PhysRevB.95.115410}%
  \BibitemOpen
  \bibfield  {author} {\bibinfo {author} {\bibfnamefont {M.~N.}\ \bibnamefont
  {Chernodub}}\ and\ \bibinfo {author} {\bibfnamefont {M.~A.}\ \bibnamefont
  {Zubkov}},\ }\bibfield  {title} {\bibinfo {title} {Chiral anomaly in {D}irac
  semimetals due to dislocations},\ }\href
  {https://doi.org/10.1103/PhysRevB.95.115410} {\bibfield  {journal} {\bibinfo
  {journal} {Phys. Rev. B}\ }\textbf {\bibinfo {volume} {95}},\ \bibinfo
  {pages} {115410} (\bibinfo {year} {2017})}\BibitemShut {NoStop}%
\bibitem [{\citenamefont {Sukhachov}\ and\ \citenamefont
  {Rostami}(2020)}]{PhysRevLett.124.126602}%
  \BibitemOpen
  \bibfield  {author} {\bibinfo {author} {\bibfnamefont {P.~O.}\ \bibnamefont
  {Sukhachov}}\ and\ \bibinfo {author} {\bibfnamefont {H.}~\bibnamefont
  {Rostami}},\ }\bibfield  {title} {\bibinfo {title} {Acoustogalvanic effect in
  {D}irac and {W}eyl semimetals},\ }\href
  {https://doi.org/10.1103/PhysRevLett.124.126602} {\bibfield  {journal}
  {\bibinfo  {journal} {Phys. Rev. Lett.}\ }\textbf {\bibinfo {volume} {124}},\
  \bibinfo {pages} {126602} (\bibinfo {year} {2020})}\BibitemShut {NoStop}%
\bibitem [{\citenamefont {Liang}\ and\ \citenamefont
  {Ojanen}(2020)}]{LLTO2020}%
  \BibitemOpen
  \bibfield  {author} {\bibinfo {author} {\bibfnamefont {L.}~\bibnamefont
  {Liang}}\ and\ \bibinfo {author} {\bibfnamefont {T.}~\bibnamefont {Ojanen}},\
  }\bibfield  {title} {\bibinfo {title} {Topological magnetotorsional effect in
  weyl semimetals},\ }\href {https://doi.org/10.1103/PhysRevResearch.2.022016}
  {\bibfield  {journal} {\bibinfo  {journal} {Phys. Rev. Res.}\ }\textbf
  {\bibinfo {volume} {2}},\ \bibinfo {pages} {022016} (\bibinfo {year}
  {2020})}\BibitemShut {NoStop}%
\bibitem [{\citenamefont {Ghosh}\ \emph {et~al.}(2020)\citenamefont {Ghosh},
  \citenamefont {Sinha}, \citenamefont {Nandy},\ and\ \citenamefont
  {Taraphder}}]{PhysRevB.102.121105}%
  \BibitemOpen
  \bibfield  {author} {\bibinfo {author} {\bibfnamefont {S.}~\bibnamefont
  {Ghosh}}, \bibinfo {author} {\bibfnamefont {D.}~\bibnamefont {Sinha}},
  \bibinfo {author} {\bibfnamefont {S.}~\bibnamefont {Nandy}},\ and\ \bibinfo
  {author} {\bibfnamefont {A.}~\bibnamefont {Taraphder}},\ }\bibfield  {title}
  {\bibinfo {title} {Chirality-dependent planar {H}all effect in inhomogeneous
  {W}eyl semimetals},\ }\href {https://doi.org/10.1103/PhysRevB.102.121105}
  {\bibfield  {journal} {\bibinfo  {journal} {Phys. Rev. B}\ }\textbf {\bibinfo
  {volume} {102}},\ \bibinfo {pages} {121105} (\bibinfo {year}
  {2020})}\BibitemShut {NoStop}%
\bibitem [{\citenamefont {Hannukainen}\ \emph {et~al.}(2020)\citenamefont
  {Hannukainen}, \citenamefont {Ferreiros}, \citenamefont {Cortijo},\ and\
  \citenamefont {Bardarson}}]{PhysRevB.102.241401}%
  \BibitemOpen
  \bibfield  {author} {\bibinfo {author} {\bibfnamefont {J.~D.}\ \bibnamefont
  {Hannukainen}}, \bibinfo {author} {\bibfnamefont {Y.}~\bibnamefont
  {Ferreiros}}, \bibinfo {author} {\bibfnamefont {A.}~\bibnamefont {Cortijo}},\
  and\ \bibinfo {author} {\bibfnamefont {J.~H.}\ \bibnamefont {Bardarson}},\
  }\bibfield  {title} {\bibinfo {title} {Axial anomaly generation by domain
  wall motion in {W}eyl semimetals},\ }\href
  {https://doi.org/10.1103/PhysRevB.102.241401} {\bibfield  {journal} {\bibinfo
   {journal} {Phys. Rev. B}\ }\textbf {\bibinfo {volume} {102}},\ \bibinfo
  {pages} {241401(R)} (\bibinfo {year} {2020})}\BibitemShut {NoStop}%
\bibitem [{\citenamefont {Liang}\ \emph {et~al.}(2021)\citenamefont {Liang},
  \citenamefont {Sukhachov},\ and\ \citenamefont {Balatsky}}]{AMEE}%
  \BibitemOpen
  \bibfield  {author} {\bibinfo {author} {\bibfnamefont {L.}~\bibnamefont
  {Liang}}, \bibinfo {author} {\bibfnamefont {P.~O.}\ \bibnamefont
  {Sukhachov}},\ and\ \bibinfo {author} {\bibfnamefont {A.~V.}\ \bibnamefont
  {Balatsky}},\ }\bibfield  {title} {\bibinfo {title} {Axial magnetoelectric
  effect in {D}irac semimetals},\ }\href
  {https://doi.org/10.1103/PhysRevLett.126.247202} {\bibfield  {journal}
  {\bibinfo  {journal} {Phys. Rev. Lett.}\ }\textbf {\bibinfo {volume} {126}},\
  \bibinfo {pages} {247202} (\bibinfo {year} {2021})}\BibitemShut {NoStop}%
\bibitem [{\citenamefont {Herasymchuk}\ \emph {et~al.}(2022)\citenamefont
  {Herasymchuk}, \citenamefont {Sukhachov},\ and\ \citenamefont
  {Gorbar}}]{PhysRevB.106.045132}%
  \BibitemOpen
  \bibfield  {author} {\bibinfo {author} {\bibfnamefont {A.~A.}\ \bibnamefont
  {Herasymchuk}}, \bibinfo {author} {\bibfnamefont {P.~O.}\ \bibnamefont
  {Sukhachov}},\ and\ \bibinfo {author} {\bibfnamefont {E.~V.}\ \bibnamefont
  {Gorbar}},\ }\bibfield  {title} {\bibinfo {title} {Electric and chiral
  response to a pseudoelectric field in {W}eyl materials},\ }\href
  {https://doi.org/10.1103/PhysRevB.106.045132} {\bibfield  {journal} {\bibinfo
   {journal} {Phys. Rev. B}\ }\textbf {\bibinfo {volume} {106}},\ \bibinfo
  {pages} {045132} (\bibinfo {year} {2022})}\BibitemShut {NoStop}%
\bibitem [{\citenamefont {Liang}(2023)}]{ACME}%
  \BibitemOpen
  \bibfield  {author} {\bibinfo {author} {\bibfnamefont {L.}~\bibnamefont
  {Liang}},\ }\bibfield  {title} {\bibinfo {title} {Anomalous chiral magnetic
  effect in time reversal symmetry breaking {W}eyl semimetals},\ }\href
  {https://doi.org/10.1103/PhysRevB.107.125101} {\bibfield  {journal} {\bibinfo
   {journal} {Phys. Rev. B}\ }\textbf {\bibinfo {volume} {107}},\ \bibinfo
  {pages} {125101} (\bibinfo {year} {2023})}\BibitemShut {NoStop}%
\bibitem [{\citenamefont {Shi}\ \emph {et~al.}(2018)\citenamefont {Shi},
  \citenamefont {Muechler}, \citenamefont {Manna}, \citenamefont {Zhang},
  \citenamefont {Koepernik}, \citenamefont {Car}, \citenamefont {van~den
  Brink}, \citenamefont {Felser},\ and\ \citenamefont {Sun}}]{YanSun1}%
  \BibitemOpen
  \bibfield  {author} {\bibinfo {author} {\bibfnamefont {W.}~\bibnamefont
  {Shi}}, \bibinfo {author} {\bibfnamefont {L.}~\bibnamefont {Muechler}},
  \bibinfo {author} {\bibfnamefont {K.}~\bibnamefont {Manna}}, \bibinfo
  {author} {\bibfnamefont {Y.}~\bibnamefont {Zhang}}, \bibinfo {author}
  {\bibfnamefont {K.}~\bibnamefont {Koepernik}}, \bibinfo {author}
  {\bibfnamefont {R.}~\bibnamefont {Car}}, \bibinfo {author} {\bibfnamefont
  {J.}~\bibnamefont {van~den Brink}}, \bibinfo {author} {\bibfnamefont
  {C.}~\bibnamefont {Felser}},\ and\ \bibinfo {author} {\bibfnamefont
  {Y.}~\bibnamefont {Sun}},\ }\bibfield  {title} {\bibinfo {title} {Prediction
  of a magnetic {W}eyl semimetal without spin-orbit coupling and strong
  anomalous {H}all effect in the {H}eusler compensated ferrimagnet
  {Ti}$_2${MnAl}},\ }\href {https://doi.org/10.1103/PhysRevB.97.060406}
  {\bibfield  {journal} {\bibinfo  {journal} {Phys. Rev. B}\ }\textbf {\bibinfo
  {volume} {97}},\ \bibinfo {pages} {060406} (\bibinfo {year}
  {2018})}\BibitemShut {NoStop}%
\bibitem [{\citenamefont {Noky}\ \emph {et~al.}(2018)\citenamefont {Noky},
  \citenamefont {Gayles}, \citenamefont {Felser},\ and\ \citenamefont
  {Sun}}]{YanSun2}%
  \BibitemOpen
  \bibfield  {author} {\bibinfo {author} {\bibfnamefont {J.}~\bibnamefont
  {Noky}}, \bibinfo {author} {\bibfnamefont {J.}~\bibnamefont {Gayles}},
  \bibinfo {author} {\bibfnamefont {C.}~\bibnamefont {Felser}},\ and\ \bibinfo
  {author} {\bibfnamefont {Y.}~\bibnamefont {Sun}},\ }\bibfield  {title}
  {\bibinfo {title} {Strong anomalous {N}ernst effect in collinear magnetic
  {W}eyl semimetals without net magnetic moments},\ }\href
  {https://doi.org/10.1103/PhysRevB.97.220405} {\bibfield  {journal} {\bibinfo
  {journal} {Phys. Rev. B}\ }\textbf {\bibinfo {volume} {97}},\ \bibinfo
  {pages} {220405} (\bibinfo {year} {2018})}\BibitemShut {NoStop}%
\bibitem [{\citenamefont {Grassano}\ \emph {et~al.}(2024)\citenamefont
  {Grassano}, \citenamefont {Binci},\ and\ \citenamefont {Marzari}}]{Heusler}%
  \BibitemOpen
  \bibfield  {author} {\bibinfo {author} {\bibfnamefont {D.}~\bibnamefont
  {Grassano}}, \bibinfo {author} {\bibfnamefont {L.}~\bibnamefont {Binci}},\
  and\ \bibinfo {author} {\bibfnamefont {N.}~\bibnamefont {Marzari}},\
  }\bibfield  {title} {\bibinfo {title} {Type-{I} antiferromagnetic {W}eyl
  semimetal {InMnTi}$_{2}$},\ }\href
  {https://doi.org/10.1103/PhysRevResearch.6.013140} {\bibfield  {journal}
  {\bibinfo  {journal} {Phys. Rev. Res.}\ }\textbf {\bibinfo {volume} {6}},\
  \bibinfo {pages} {013140} (\bibinfo {year} {2024})}\BibitemShut {NoStop}%
\bibitem [{\citenamefont {Chang}\ \emph {et~al.}(2018)\citenamefont {Chang},
  \citenamefont {Singh}, \citenamefont {Xu}, \citenamefont {Bian},
  \citenamefont {Huang}, \citenamefont {Hsu}, \citenamefont {Belopolski},
  \citenamefont {Alidoust}, \citenamefont {Sanchez}, \citenamefont {Zheng},
  \citenamefont {Lu}, \citenamefont {Zhang}, \citenamefont {Bian},
  \citenamefont {Chang}, \citenamefont {Jeng}, \citenamefont {Bansil},
  \citenamefont {Hsu}, \citenamefont {Jia}, \citenamefont {Neupert},
  \citenamefont {Lin},\ and\ \citenamefont {Hasan}}]{PhysRevB.97.041104}%
  \BibitemOpen
  \bibfield  {author} {\bibinfo {author} {\bibfnamefont {G.}~\bibnamefont
  {Chang}}, \bibinfo {author} {\bibfnamefont {B.}~\bibnamefont {Singh}},
  \bibinfo {author} {\bibfnamefont {S.-Y.}\ \bibnamefont {Xu}}, \bibinfo
  {author} {\bibfnamefont {G.}~\bibnamefont {Bian}}, \bibinfo {author}
  {\bibfnamefont {S.-M.}\ \bibnamefont {Huang}}, \bibinfo {author}
  {\bibfnamefont {C.-H.}\ \bibnamefont {Hsu}}, \bibinfo {author} {\bibfnamefont
  {I.}~\bibnamefont {Belopolski}}, \bibinfo {author} {\bibfnamefont
  {N.}~\bibnamefont {Alidoust}}, \bibinfo {author} {\bibfnamefont {D.~S.}\
  \bibnamefont {Sanchez}}, \bibinfo {author} {\bibfnamefont {H.}~\bibnamefont
  {Zheng}}, \bibinfo {author} {\bibfnamefont {H.}~\bibnamefont {Lu}}, \bibinfo
  {author} {\bibfnamefont {X.}~\bibnamefont {Zhang}}, \bibinfo {author}
  {\bibfnamefont {Y.}~\bibnamefont {Bian}}, \bibinfo {author} {\bibfnamefont
  {T.-R.}\ \bibnamefont {Chang}}, \bibinfo {author} {\bibfnamefont {H.-T.}\
  \bibnamefont {Jeng}}, \bibinfo {author} {\bibfnamefont {A.}~\bibnamefont
  {Bansil}}, \bibinfo {author} {\bibfnamefont {H.}~\bibnamefont {Hsu}},
  \bibinfo {author} {\bibfnamefont {S.}~\bibnamefont {Jia}}, \bibinfo {author}
  {\bibfnamefont {T.}~\bibnamefont {Neupert}}, \bibinfo {author} {\bibfnamefont
  {H.}~\bibnamefont {Lin}},\ and\ \bibinfo {author} {\bibfnamefont {M.~Z.}\
  \bibnamefont {Hasan}},\ }\bibfield  {title} {\bibinfo {title} {Magnetic and
  noncentrosymmetric {W}eyl fermion semimetals in the {RAlGe} family of
  compounds ({R}=rare earth)},\ }\href
  {https://doi.org/10.1103/PhysRevB.97.041104} {\bibfield  {journal} {\bibinfo
  {journal} {Phys. Rev. B}\ }\textbf {\bibinfo {volume} {97}},\ \bibinfo
  {pages} {041104} (\bibinfo {year} {2018})}\BibitemShut {NoStop}%
\bibitem [{\citenamefont {Sanchez}\ \emph {et~al.}(2020)\citenamefont
  {Sanchez}, \citenamefont {Chang}, \citenamefont {Belopolski}, \citenamefont
  {Lu}, \citenamefont {Yin}, \citenamefont {Alidoust}, \citenamefont {Xu},
  \citenamefont {Cochran}, \citenamefont {Zhang}, \citenamefont {Bian},
  \citenamefont {Zhang}, \citenamefont {Liu}, \citenamefont {Ma}, \citenamefont
  {Bian}, \citenamefont {Lin}, \citenamefont {Xu}, \citenamefont {Jia},\ and\
  \citenamefont {Hasan}}]{Sanchez2020}%
  \BibitemOpen
  \bibfield  {author} {\bibinfo {author} {\bibfnamefont {D.~S.}\ \bibnamefont
  {Sanchez}}, \bibinfo {author} {\bibfnamefont {G.}~\bibnamefont {Chang}},
  \bibinfo {author} {\bibfnamefont {I.}~\bibnamefont {Belopolski}}, \bibinfo
  {author} {\bibfnamefont {H.}~\bibnamefont {Lu}}, \bibinfo {author}
  {\bibfnamefont {J.-X.}\ \bibnamefont {Yin}}, \bibinfo {author} {\bibfnamefont
  {N.}~\bibnamefont {Alidoust}}, \bibinfo {author} {\bibfnamefont
  {X.}~\bibnamefont {Xu}}, \bibinfo {author} {\bibfnamefont {T.~A.}\
  \bibnamefont {Cochran}}, \bibinfo {author} {\bibfnamefont {X.}~\bibnamefont
  {Zhang}}, \bibinfo {author} {\bibfnamefont {Y.}~\bibnamefont {Bian}},
  \bibinfo {author} {\bibfnamefont {S.~S.}\ \bibnamefont {Zhang}}, \bibinfo
  {author} {\bibfnamefont {Y.-Y.}\ \bibnamefont {Liu}}, \bibinfo {author}
  {\bibfnamefont {J.}~\bibnamefont {Ma}}, \bibinfo {author} {\bibfnamefont
  {G.}~\bibnamefont {Bian}}, \bibinfo {author} {\bibfnamefont {H.}~\bibnamefont
  {Lin}}, \bibinfo {author} {\bibfnamefont {S.-Y.}\ \bibnamefont {Xu}},
  \bibinfo {author} {\bibfnamefont {S.}~\bibnamefont {Jia}},\ and\ \bibinfo
  {author} {\bibfnamefont {M.~Z.}\ \bibnamefont {Hasan}},\ }\bibfield  {title}
  {\bibinfo {title} {Observation of {W}eyl fermions in a magnetic
  non-centrosymmetric crystal},\ }\href
  {https://doi.org/10.1038/s41467-020-16879-1} {\bibfield  {journal} {\bibinfo
  {journal} {Nat. Commun.}\ }\textbf {\bibinfo {volume} {11}},\ \bibinfo
  {pages} {3356} (\bibinfo {year} {2020})}\BibitemShut {NoStop}%
\bibitem [{\citenamefont {Gaudet}\ \emph {et~al.}(2021)\citenamefont {Gaudet},
  \citenamefont {Yang}, \citenamefont {Baidya}, \citenamefont {Lu},
  \citenamefont {Xu}, \citenamefont {Zhao}, \citenamefont {Rodriguez-Rivera},
  \citenamefont {Hoffmann}, \citenamefont {Graf}, \citenamefont {Torchinsky},
  \citenamefont {Nikoli{\'{c}}}, \citenamefont {Vanderbilt}, \citenamefont
  {Tafti},\ and\ \citenamefont {Broholm}}]{Gaudet2021}%
  \BibitemOpen
  \bibfield  {author} {\bibinfo {author} {\bibfnamefont {J.}~\bibnamefont
  {Gaudet}}, \bibinfo {author} {\bibfnamefont {H.-Y.}\ \bibnamefont {Yang}},
  \bibinfo {author} {\bibfnamefont {S.}~\bibnamefont {Baidya}}, \bibinfo
  {author} {\bibfnamefont {B.}~\bibnamefont {Lu}}, \bibinfo {author}
  {\bibfnamefont {G.}~\bibnamefont {Xu}}, \bibinfo {author} {\bibfnamefont
  {Y.}~\bibnamefont {Zhao}}, \bibinfo {author} {\bibfnamefont {J.~A.}\
  \bibnamefont {Rodriguez-Rivera}}, \bibinfo {author} {\bibfnamefont {C.~M.}\
  \bibnamefont {Hoffmann}}, \bibinfo {author} {\bibfnamefont {D.~E.}\
  \bibnamefont {Graf}}, \bibinfo {author} {\bibfnamefont {D.~H.}\ \bibnamefont
  {Torchinsky}}, \bibinfo {author} {\bibfnamefont {P.}~\bibnamefont
  {Nikoli{\'{c}}}}, \bibinfo {author} {\bibfnamefont {D.}~\bibnamefont
  {Vanderbilt}}, \bibinfo {author} {\bibfnamefont {F.}~\bibnamefont {Tafti}},\
  and\ \bibinfo {author} {\bibfnamefont {C.~L.}\ \bibnamefont {Broholm}},\
  }\bibfield  {title} {\bibinfo {title} {Weyl-mediated helical magnetism in
  {NdAlSi}},\ }\href {https://doi.org/10.1038/s41563-021-01062-8} {\bibfield
  {journal} {\bibinfo  {journal} {Nat. Mater.}\ }\textbf {\bibinfo {volume}
  {20}},\ \bibinfo {pages} {1650} (\bibinfo {year} {2021})}\BibitemShut
  {NoStop}%
\bibitem [{\citenamefont {Yao}\ \emph {et~al.}(2023)\citenamefont {Yao},
  \citenamefont {Gaudet}, \citenamefont {Verma}, \citenamefont {Graf},
  \citenamefont {Yang}, \citenamefont {Bahrami}, \citenamefont {Zhang},
  \citenamefont {Aczel}, \citenamefont {Subedi}, \citenamefont {Torchinsky},
  \citenamefont {Sun}, \citenamefont {Bansil}, \citenamefont {Huang},
  \citenamefont {Singh}, \citenamefont {Blaha}, \citenamefont
  {Nikoli\ifmmode~\acute{c}\else \'{c}\fi{}},\ and\ \citenamefont
  {Tafti}}]{PhysRevX.13.011035}%
  \BibitemOpen
  \bibfield  {author} {\bibinfo {author} {\bibfnamefont {X.}~\bibnamefont
  {Yao}}, \bibinfo {author} {\bibfnamefont {J.}~\bibnamefont {Gaudet}},
  \bibinfo {author} {\bibfnamefont {R.}~\bibnamefont {Verma}}, \bibinfo
  {author} {\bibfnamefont {D.~E.}\ \bibnamefont {Graf}}, \bibinfo {author}
  {\bibfnamefont {H.-Y.}\ \bibnamefont {Yang}}, \bibinfo {author}
  {\bibfnamefont {F.}~\bibnamefont {Bahrami}}, \bibinfo {author} {\bibfnamefont
  {R.}~\bibnamefont {Zhang}}, \bibinfo {author} {\bibfnamefont {A.~A.}\
  \bibnamefont {Aczel}}, \bibinfo {author} {\bibfnamefont {S.}~\bibnamefont
  {Subedi}}, \bibinfo {author} {\bibfnamefont {D.~H.}\ \bibnamefont
  {Torchinsky}}, \bibinfo {author} {\bibfnamefont {J.}~\bibnamefont {Sun}},
  \bibinfo {author} {\bibfnamefont {A.}~\bibnamefont {Bansil}}, \bibinfo
  {author} {\bibfnamefont {S.-M.}\ \bibnamefont {Huang}}, \bibinfo {author}
  {\bibfnamefont {B.}~\bibnamefont {Singh}}, \bibinfo {author} {\bibfnamefont
  {P.}~\bibnamefont {Blaha}}, \bibinfo {author} {\bibfnamefont
  {P.}~\bibnamefont {Nikoli\ifmmode~\acute{c}\else \'{c}\fi{}}},\ and\ \bibinfo
  {author} {\bibfnamefont {F.}~\bibnamefont {Tafti}},\ }\bibfield  {title}
  {\bibinfo {title} {Large topological {H}all effect and spiral magnetic order
  in the {W}eyl semimetal {SmAlSi}},\ }\href
  {https://doi.org/10.1103/PhysRevX.13.011035} {\bibfield  {journal} {\bibinfo
  {journal} {Phys. Rev. X}\ }\textbf {\bibinfo {volume} {13}},\ \bibinfo
  {pages} {011035} (\bibinfo {year} {2023})}\BibitemShut {NoStop}%
\bibitem [{\citenamefont {Drucker}\ \emph {et~al.}(2023)\citenamefont
  {Drucker}, \citenamefont {Nguyen}, \citenamefont {Han}, \citenamefont
  {Siriviboon}, \citenamefont {Luo}, \citenamefont {Andrejevic}, \citenamefont
  {Zhu}, \citenamefont {Bednik}, \citenamefont {Nguyen}, \citenamefont {Chen},
  \citenamefont {Nguyen}, \citenamefont {Liu}, \citenamefont {Williams},
  \citenamefont {Stone}, \citenamefont {Kolesnikov}, \citenamefont {Chi},
  \citenamefont {Fernandez-Baca}, \citenamefont {Nelson}, \citenamefont
  {Alatas}, \citenamefont {Hogan}, \citenamefont {Puretzky}, \citenamefont
  {Huang}, \citenamefont {Yu},\ and\ \citenamefont {Li}}]{Drucker2023}%
  \BibitemOpen
  \bibfield  {author} {\bibinfo {author} {\bibfnamefont {N.~C.}\ \bibnamefont
  {Drucker}}, \bibinfo {author} {\bibfnamefont {T.}~\bibnamefont {Nguyen}},
  \bibinfo {author} {\bibfnamefont {F.}~\bibnamefont {Han}}, \bibinfo {author}
  {\bibfnamefont {P.}~\bibnamefont {Siriviboon}}, \bibinfo {author}
  {\bibfnamefont {X.}~\bibnamefont {Luo}}, \bibinfo {author} {\bibfnamefont
  {N.}~\bibnamefont {Andrejevic}}, \bibinfo {author} {\bibfnamefont
  {Z.}~\bibnamefont {Zhu}}, \bibinfo {author} {\bibfnamefont {G.}~\bibnamefont
  {Bednik}}, \bibinfo {author} {\bibfnamefont {Q.~T.}\ \bibnamefont {Nguyen}},
  \bibinfo {author} {\bibfnamefont {Z.}~\bibnamefont {Chen}}, \bibinfo {author}
  {\bibfnamefont {L.~K.}\ \bibnamefont {Nguyen}}, \bibinfo {author}
  {\bibfnamefont {T.}~\bibnamefont {Liu}}, \bibinfo {author} {\bibfnamefont
  {T.~J.}\ \bibnamefont {Williams}}, \bibinfo {author} {\bibfnamefont {M.~B.}\
  \bibnamefont {Stone}}, \bibinfo {author} {\bibfnamefont {A.~I.}\ \bibnamefont
  {Kolesnikov}}, \bibinfo {author} {\bibfnamefont {S.}~\bibnamefont {Chi}},
  \bibinfo {author} {\bibfnamefont {J.}~\bibnamefont {Fernandez-Baca}},
  \bibinfo {author} {\bibfnamefont {C.~S.}\ \bibnamefont {Nelson}}, \bibinfo
  {author} {\bibfnamefont {A.}~\bibnamefont {Alatas}}, \bibinfo {author}
  {\bibfnamefont {T.}~\bibnamefont {Hogan}}, \bibinfo {author} {\bibfnamefont
  {A.~A.}\ \bibnamefont {Puretzky}}, \bibinfo {author} {\bibfnamefont
  {S.}~\bibnamefont {Huang}}, \bibinfo {author} {\bibfnamefont
  {Y.}~\bibnamefont {Yu}},\ and\ \bibinfo {author} {\bibfnamefont
  {M.}~\bibnamefont {Li}},\ }\bibfield  {title} {\bibinfo {title} {Topology
  stabilized fluctuations in a magnetic nodal semimetal},\ }\href
  {https://doi.org/10.1038/s41467-023-40765-1} {\bibfield  {journal} {\bibinfo
  {journal} {Nat. Commun.}\ }\textbf {\bibinfo {volume} {14}},\ \bibinfo
  {pages} {5182} (\bibinfo {year} {2023})}\BibitemShut {NoStop}%
\bibitem [{\citenamefont {Laha}\ \emph {et~al.}(2024)\citenamefont {Laha},
  \citenamefont {Kundu}, \citenamefont {Aryal}, \citenamefont {Bozin},
  \citenamefont {Yao}, \citenamefont {Paone}, \citenamefont {Rajapitamahuni},
  \citenamefont {Vescovo}, \citenamefont {Valla}, \citenamefont {Abeykoon},
  \citenamefont {Jing}, \citenamefont {Yin}, \citenamefont {Pasupathy},
  \citenamefont {Liu},\ and\ \citenamefont {Li}}]{PhysRevB.109.035120}%
  \BibitemOpen
  \bibfield  {author} {\bibinfo {author} {\bibfnamefont {A.}~\bibnamefont
  {Laha}}, \bibinfo {author} {\bibfnamefont {A.~K.}\ \bibnamefont {Kundu}},
  \bibinfo {author} {\bibfnamefont {N.}~\bibnamefont {Aryal}}, \bibinfo
  {author} {\bibfnamefont {E.~S.}\ \bibnamefont {Bozin}}, \bibinfo {author}
  {\bibfnamefont {J.}~\bibnamefont {Yao}}, \bibinfo {author} {\bibfnamefont
  {S.}~\bibnamefont {Paone}}, \bibinfo {author} {\bibfnamefont
  {A.}~\bibnamefont {Rajapitamahuni}}, \bibinfo {author} {\bibfnamefont
  {E.}~\bibnamefont {Vescovo}}, \bibinfo {author} {\bibfnamefont
  {T.}~\bibnamefont {Valla}}, \bibinfo {author} {\bibfnamefont
  {M.}~\bibnamefont {Abeykoon}}, \bibinfo {author} {\bibfnamefont
  {R.}~\bibnamefont {Jing}}, \bibinfo {author} {\bibfnamefont {W.}~\bibnamefont
  {Yin}}, \bibinfo {author} {\bibfnamefont {A.~N.}\ \bibnamefont {Pasupathy}},
  \bibinfo {author} {\bibfnamefont {M.}~\bibnamefont {Liu}},\ and\ \bibinfo
  {author} {\bibfnamefont {Q.}~\bibnamefont {Li}},\ }\bibfield  {title}
  {\bibinfo {title} {{Electronic structure and magnetic and transport
  properties of antiferromagnetic Weyl semimetal GdAlSi}},\ }\href
  {https://doi.org/10.1103/PhysRevB.109.035120} {\bibfield  {journal} {\bibinfo
   {journal} {Phys. Rev. B}\ }\textbf {\bibinfo {volume} {109}},\ \bibinfo
  {pages} {035120} (\bibinfo {year} {2024})}\BibitemShut {NoStop}%
\bibitem [{\citenamefont {Kunze}\ \emph {et~al.}(2024)\citenamefont {Kunze},
  \citenamefont {K\"opf}, \citenamefont {Cao}, \citenamefont {Qi},\ and\
  \citenamefont {Kuntscher}}]{PhysRevB.109.195130}%
  \BibitemOpen
  \bibfield  {author} {\bibinfo {author} {\bibfnamefont {J.}~\bibnamefont
  {Kunze}}, \bibinfo {author} {\bibfnamefont {M.}~\bibnamefont {K\"opf}},
  \bibinfo {author} {\bibfnamefont {W.}~\bibnamefont {Cao}}, \bibinfo {author}
  {\bibfnamefont {Y.}~\bibnamefont {Qi}},\ and\ \bibinfo {author}
  {\bibfnamefont {C.~A.}\ \bibnamefont {Kuntscher}},\ }\bibfield  {title}
  {\bibinfo {title} {Optical signatures of type-{II} {W}eyl fermions in the
  noncentrosymmetric semimetals {RAlSi} ({R}={La}, {Ce}, {Pr}, {Nd}, {Sm})},\
  }\href {https://doi.org/10.1103/PhysRevB.109.195130} {\bibfield  {journal}
  {\bibinfo  {journal} {Phys. Rev. B}\ }\textbf {\bibinfo {volume} {109}},\
  \bibinfo {pages} {195130} (\bibinfo {year} {2024})}\BibitemShut {NoStop}%
\bibitem [{\citenamefont {Varjas}\ \emph {et~al.}(2016)\citenamefont {Varjas},
  \citenamefont {Grushin}, \citenamefont {Ilan},\ and\ \citenamefont
  {Moore}}]{PhysRevLett.117.257601}%
  \BibitemOpen
  \bibfield  {author} {\bibinfo {author} {\bibfnamefont {D.}~\bibnamefont
  {Varjas}}, \bibinfo {author} {\bibfnamefont {A.~G.}\ \bibnamefont {Grushin}},
  \bibinfo {author} {\bibfnamefont {R.}~\bibnamefont {Ilan}},\ and\ \bibinfo
  {author} {\bibfnamefont {J.~E.}\ \bibnamefont {Moore}},\ }\bibfield  {title}
  {\bibinfo {title} {Dynamical piezoelectric and magnetopiezoelectric effects
  in polar metals from {B}erry phases and orbital moments},\ }\href
  {https://doi.org/10.1103/PhysRevLett.117.257601} {\bibfield  {journal}
  {\bibinfo  {journal} {Phys. Rev. Lett.}\ }\textbf {\bibinfo {volume} {117}},\
  \bibinfo {pages} {257601} (\bibinfo {year} {2016})}\BibitemShut {NoStop}%
\bibitem [{\citenamefont {Watanabe}\ and\ \citenamefont
  {Yanase}(2017)}]{PhysRevB.96.064432}%
  \BibitemOpen
  \bibfield  {author} {\bibinfo {author} {\bibfnamefont {H.}~\bibnamefont
  {Watanabe}}\ and\ \bibinfo {author} {\bibfnamefont {Y.}~\bibnamefont
  {Yanase}},\ }\bibfield  {title} {\bibinfo {title} {Magnetic hexadecapole
  order and magnetopiezoelectric metal state in
  {Ba}$_{1-x}${K}$_{x}${Mn}$_{2}${As}$_{2}$},\ }\href
  {https://doi.org/10.1103/PhysRevB.96.064432} {\bibfield  {journal} {\bibinfo
  {journal} {Phys. Rev. B}\ }\textbf {\bibinfo {volume} {96}},\ \bibinfo
  {pages} {064432} (\bibinfo {year} {2017})}\BibitemShut {NoStop}%
\bibitem [{\citenamefont {Rodriguez-Lopez}\ and\ \citenamefont
  {Cortijo}(2018)}]{PhysRevB.97.235128}%
  \BibitemOpen
  \bibfield  {author} {\bibinfo {author} {\bibfnamefont {P.}~\bibnamefont
  {Rodriguez-Lopez}}\ and\ \bibinfo {author} {\bibfnamefont {A.}~\bibnamefont
  {Cortijo}},\ }\bibfield  {title} {\bibinfo {title} {Theory of the
  strain-induced magnetoelectric effect in planar {D}irac systems},\ }\href
  {https://doi.org/10.1103/PhysRevB.97.235128} {\bibfield  {journal} {\bibinfo
  {journal} {Phys. Rev. B}\ }\textbf {\bibinfo {volume} {97}},\ \bibinfo
  {pages} {235128} (\bibinfo {year} {2018})}\BibitemShut {NoStop}%
\bibitem [{\citenamefont {Shiomi}\ \emph {et~al.}(2019)\citenamefont {Shiomi},
  \citenamefont {Watanabe}, \citenamefont {Masuda}, \citenamefont {Takahashi},
  \citenamefont {Yanase},\ and\ \citenamefont
  {Ishiwata}}]{PhysRevLett.122.127207}%
  \BibitemOpen
  \bibfield  {author} {\bibinfo {author} {\bibfnamefont {Y.}~\bibnamefont
  {Shiomi}}, \bibinfo {author} {\bibfnamefont {H.}~\bibnamefont {Watanabe}},
  \bibinfo {author} {\bibfnamefont {H.}~\bibnamefont {Masuda}}, \bibinfo
  {author} {\bibfnamefont {H.}~\bibnamefont {Takahashi}}, \bibinfo {author}
  {\bibfnamefont {Y.}~\bibnamefont {Yanase}},\ and\ \bibinfo {author}
  {\bibfnamefont {S.}~\bibnamefont {Ishiwata}},\ }\bibfield  {title} {\bibinfo
  {title} {Observation of a magnetopiezoelectric effect in the
  antiferromagnetic metal {EuMnBi}$_{2}$},\ }\href
  {https://doi.org/10.1103/PhysRevLett.122.127207} {\bibfield  {journal}
  {\bibinfo  {journal} {Phys. Rev. Lett.}\ }\textbf {\bibinfo {volume} {122}},\
  \bibinfo {pages} {127207} (\bibinfo {year} {2019})}\BibitemShut {NoStop}%
\bibitem [{\citenamefont {Belitz}\ \emph {et~al.}(2006)\citenamefont {Belitz},
  \citenamefont {Kirkpatrick},\ and\ \citenamefont
  {Rosch}}]{PhysRevB.73.054431}%
  \BibitemOpen
  \bibfield  {author} {\bibinfo {author} {\bibfnamefont {D.}~\bibnamefont
  {Belitz}}, \bibinfo {author} {\bibfnamefont {T.~R.}\ \bibnamefont
  {Kirkpatrick}},\ and\ \bibinfo {author} {\bibfnamefont {A.}~\bibnamefont
  {Rosch}},\ }\bibfield  {title} {\bibinfo {title} {Theory of helimagnons in
  itinerant quantum systems},\ }\href
  {https://doi.org/10.1103/PhysRevB.73.054431} {\bibfield  {journal} {\bibinfo
  {journal} {Phys. Rev. B}\ }\textbf {\bibinfo {volume} {73}},\ \bibinfo
  {pages} {054431} (\bibinfo {year} {2006})}\BibitemShut {NoStop}%
\bibitem [{Note1()}]{Note1}%
  \BibitemOpen
  \bibinfo {note} {The Fermi velocity $v_F$ of Weyl semimetals is of the order
  of $10^6$\si {m/s}, while the sound velocity $v_s$ is of the order of
  $10^3$\si {m/s}, so $\alpha =v_s/v_F\ll 1$. The spin stiffness $D$ is
  typically of the order of \si {meV\r A}$^2$, while $\hbar v_F$ is of the
  order of \si {eV\r A} and the Fermi momentum $k_F$ is of the order of \si {\r
  A}$^{-1}$, thus $\gamma =D k_F/v_F\ll 1$.}\BibitemShut {Stop}%
\bibitem [{\citenamefont {Altland}\ and\ \citenamefont
  {Simons}(2010)}]{altland_simons_2010}%
  \BibitemOpen
  \bibfield  {author} {\bibinfo {author} {\bibfnamefont {A.}~\bibnamefont
  {Altland}}\ and\ \bibinfo {author} {\bibfnamefont {B.~D.}\ \bibnamefont
  {Simons}},\ }\href {https://doi.org/10.1017/CBO9780511789984} {\emph
  {\bibinfo {title} {Condensed Matter Field Theory}}},\ \bibinfo {edition}
  {2nd}\ ed.\ (\bibinfo  {publisher} {Cambridge University Press},\ \bibinfo
  {year} {2010})\BibitemShut {NoStop}%
\bibitem [{\citenamefont {Cortijo}\ \emph {et~al.}(2015)\citenamefont
  {Cortijo}, \citenamefont {Ferreir\'os}, \citenamefont {Landsteiner},\ and\
  \citenamefont {Vozmediano}}]{PhysRevLett.115.177202}%
  \BibitemOpen
  \bibfield  {author} {\bibinfo {author} {\bibfnamefont {A.}~\bibnamefont
  {Cortijo}}, \bibinfo {author} {\bibfnamefont {Y.}~\bibnamefont
  {Ferreir\'os}}, \bibinfo {author} {\bibfnamefont {K.}~\bibnamefont
  {Landsteiner}},\ and\ \bibinfo {author} {\bibfnamefont {M.~A.~H.}\
  \bibnamefont {Vozmediano}},\ }\bibfield  {title} {\bibinfo {title} {Elastic
  gauge fields in {W}eyl semimetals},\ }\href
  {https://doi.org/10.1103/PhysRevLett.115.177202} {\bibfield  {journal}
  {\bibinfo  {journal} {Phys. Rev. Lett.}\ }\textbf {\bibinfo {volume} {115}},\
  \bibinfo {pages} {177202} (\bibinfo {year} {2015})}\BibitemShut {NoStop}%
\bibitem [{\citenamefont {Chumak}\ \emph {et~al.}(2015)\citenamefont {Chumak},
  \citenamefont {Vasyuchka}, \citenamefont {Serga},\ and\ \citenamefont
  {Hillebrands}}]{Chumak2015}%
  \BibitemOpen
  \bibfield  {author} {\bibinfo {author} {\bibfnamefont {A.~V.}\ \bibnamefont
  {Chumak}}, \bibinfo {author} {\bibfnamefont {V.~I.}\ \bibnamefont
  {Vasyuchka}}, \bibinfo {author} {\bibfnamefont {A.~A.}\ \bibnamefont
  {Serga}},\ and\ \bibinfo {author} {\bibfnamefont {B.}~\bibnamefont
  {Hillebrands}},\ }\bibfield  {title} {\bibinfo {title} {Magnon spintronics},\
  }\href {https://doi.org/10.1038/nphys3347} {\bibfield  {journal} {\bibinfo
  {journal} {Nat. Phys.}\ }\textbf {\bibinfo {volume} {11}},\ \bibinfo {pages}
  {453} (\bibinfo {year} {2015})}\BibitemShut {NoStop}%
\bibitem [{\citenamefont {Chang}\ \emph {et~al.}(2015)\citenamefont {Chang},
  \citenamefont {Zhou}, \citenamefont {Wang}, \citenamefont {Shan},\ and\
  \citenamefont {Xiao}}]{RKKYZhang}%
  \BibitemOpen
  \bibfield  {author} {\bibinfo {author} {\bibfnamefont {H.-R.}\ \bibnamefont
  {Chang}}, \bibinfo {author} {\bibfnamefont {J.}~\bibnamefont {Zhou}},
  \bibinfo {author} {\bibfnamefont {S.-X.}\ \bibnamefont {Wang}}, \bibinfo
  {author} {\bibfnamefont {W.-Y.}\ \bibnamefont {Shan}},\ and\ \bibinfo
  {author} {\bibfnamefont {D.}~\bibnamefont {Xiao}},\ }\bibfield  {title}
  {\bibinfo {title} {{RKKY} interaction of magnetic impurities in {D}irac and
  {W}eyl semimetals},\ }\href {https://doi.org/10.1103/PhysRevB.92.241103}
  {\bibfield  {journal} {\bibinfo  {journal} {Phys. Rev. B}\ }\textbf {\bibinfo
  {volume} {92}},\ \bibinfo {pages} {241103} (\bibinfo {year}
  {2015})}\BibitemShut {NoStop}%
\bibitem [{\citenamefont {Burkov}\ \emph {et~al.}(2011)\citenamefont {Burkov},
  \citenamefont {Hook},\ and\ \citenamefont {Balents}}]{PhysRevB.84.235126}%
  \BibitemOpen
  \bibfield  {author} {\bibinfo {author} {\bibfnamefont {A.~A.}\ \bibnamefont
  {Burkov}}, \bibinfo {author} {\bibfnamefont {M.~D.}\ \bibnamefont {Hook}},\
  and\ \bibinfo {author} {\bibfnamefont {L.}~\bibnamefont {Balents}},\
  }\bibfield  {title} {\bibinfo {title} {Topological nodal semimetals},\ }\href
  {https://doi.org/10.1103/PhysRevB.84.235126} {\bibfield  {journal} {\bibinfo
  {journal} {Phys. Rev. B}\ }\textbf {\bibinfo {volume} {84}},\ \bibinfo
  {pages} {235126} (\bibinfo {year} {2011})}\BibitemShut {NoStop}%
\end{thebibliography}%
\end{document}